# Verifying the HALE measures of the Global Burden of Disease Study: Quantitative Methods Proposed*

**Christos H Skiadas**

ManLab, Technical University of Crete, Chania, Crete, Greece
Email: skiadas@cmsim.net

**Abstract**

To verify the Global Burden of Disease Study and the provided healthy life expectancy (HALE) estimates from the World Health Organization (WHO) we propose a very simple model based on the mortality μx of a population provided in a classical life table and a mortality diagram. We use the abridged life tables provided by WHO. Our estimates are compared with the HALE estimates for the World territories and the WHO countries. Even more we have developed the related simple program in Excel which provides immediately the Life Expectancy, the Loss of Healthy Life Years and the Healthy Life Expectancy estimate. We also apply the health state function theory to have more estimates and comparisons. The results suggest improved WHO estimates in recent years for the majority of the cases.

Introduction

Starting from the late 80's a Global Burden of Disease (GBD) study was applied in many countries reflecting the optimistic views of many researchers and policy makers worldwide to quantify the health state of a population or a group of persons. In the time course they succeeded in establishing an international network collecting and providing adequate information to calculate health measures under terms as Loss of Healthy Life Years (LHLY) or Healthy Life Expectancy (HALE). The latter tends to be a serious measure important for the policy makers and national and international health programs. So far the process followed was towards statistical measures including surveys and data collection using questionnaires and disability and epidemiological data as well. They faced many views referring to the definition of health and to the inability to count the various health states and of course the different cultural and societal aspects of the estimation of health by various persons worldwide. Further to any objections posed when trying to quantify health, the scientific community had simply to express with strong and reliable measures that millions of people for centuries and thousands of years expressed and continue to repeat every day: That their health is good, fair, bad or very bad. As for many decades the public opinion is seriously quantified by using well established statistical and poll techniques it is not surprising that a part of these achievements helped to improve, establish and disseminate the health state measures. However, a serious scientific part is missing or it is not very much explored that is to find the model underlying the health state measures. Observing the health state measures by country from 1990 until nowadays it is clear that the observed and estimated health parameters follow a rather

*Paper version sent to ArXiv.org (26 October 2015)

systematic way. If so why not to find the process underlying these measures? It will support the provided health measures with enough documentation while new horizons will open towards better estimates and data validation.

From the early 90's we have introduced and applied methods, models and techniques to estimate the health state of a population. The related results appear in several publications and we have already observed that our estimates are related or closely related to the provided by the World Health Organization (WHO) and other agencies as Eurostat or experts as the REVES group. However, our method based on a difficult stochastic analysis technique, is not easy to use especially by practitioners. The last four centuries demography and demographers are based on the classical Life Tables. Thus here we propose a very simple model based on the mortality $μ_x$ of a population provided in a classical life table. To compare our results with those provided by WHO we use the μx included in the WHO abridged life tables. Our estimates are compared with the HALE estimates for all the WHO countries. Even more we provide the related simple program in Excel which provides immediately the Life Expectancy, the Loss of Healthy Life Years and the Healthy Life Expectancy estimate. The comparisons suggest an improved WHO estimate for the majority of the countries. There are countries' results differing from the model and need further study.

**More Details**

The Global Burden of Disease Study explored the health status of the population of all the countries members of the World Health Organization (WHO). It is a large team work started more than 25 years ago (see Murray and Lopez, 1997,2000, Mathers et al., 2000, Salomon, et al., 2010, 2012, Murray et al., 2015, Hausman, 2012, Vos et al., 2012, WHO, 2000, 2001, 2002, 2004, 2013, 2014 and many other publications). The last years, with the financial support of the Bill and Melinda Gates foundation, the work was expanded via a large international group of researchers. The accuracy of the data collection methods was improved along with the data development and application techniques. So far the health status indicators were developed and gradually were established under terms as healthy life expectancy and loss of healthy life years. Methods and techniques developed during the seventies and eighties as the Sullivan method (Sullivan, 1971) were used quite successfully. Several publications are done with the most important included in The Lancet under the terms DALE and HALE whereas a considerable number can be found in the WHO and World Bank publications. The same half part of a century several works appear in the European Union exploring the same phenomenon and providing more insight to the estimation of the health state of a population and providing tools for the estimation of severe, moderate and light disability. The use of these estimates from the health systems and the governments is obvious.

To a surprise the development of the theoretical tools was not so large. The main direction was towards to surveys and collection of mass health state data instead of developing and using theoretical tools. The lessons learned during the last centuries were towards the introduction of models in the analysis of health and mortality. The classical examples are Edmund Halley for Life Tables and Benjamin Gompertz for the law of mortality and may others. Today our ability to use mass storage tools as the computers and the extensive application of surveys and polls to many political, social and economic activities directed the main health state studies. In other words we give much attention to opinions of the people for their health status followed by extensive health data collection. However, it remains a serious question: can we validate the health status results? As it is the standard procedure in science a systematic study as the Global Burden of Disease should be validated by one or more models. Especially as these studies are today the main tool for the health programs of many countries the need of verification is more important.

People reply according to their experience. Two main approaches arise: The mortality focus approach and the health status approach. Although both look similar responds may have significant differences. The main reason is that health is a rather optimistic word opposed to the pessimistic mortality term. Twenty years ago we provided a model to express the health state of a population. We developed and expanded this model leading to a system providing health status indexes. Here we propose a simple but yet powerful model to estimate the health indexes provided by WHO.

**The mortality approach: A simple method**

In this case the μx diagram provides a simple but quite useful estimate for the loss of healthy life years of a population. $M_x$ is provided by the related estimates of the bureau of the census of a country, the Eurostat, the World Health Organization, the Human Mortality Database and other institutions.

The way people assess disability has also to do with the information collected from the close environment (relatives, friends, office staff) and the far environment mainly communicated by the mass media, internet and other information sources.

The simplest way to have an estimate for this information is to ask people directly which of course has a large degree of uncertainty and it is subject to errors and misunderstandings due to many factors concerning human communication. An alternative is to count this information by a sort of summation as is the sum of mortality in the time course. This can be done by estimating the total influence regarding mortality in the opinions of a population as a sum of $\mu_x$ in an age interval leading to an integral in the limit as:

$$E_x = \lim_{n \to \infty} \sum_{1}^{n} \mu_x = \int_0^s \mu_s \, ds$$

Where Ex expresses the area OCABO in the mortality diagram presented in Figure 1.

The classical approach for μx is to assume a Gompertz like formula of the form (where a and b are parameters):

$$\mu_x = ae^{bx}$$

Then the resulting value for Ex in the interval [0, T] is:

$$E_x = \frac{a}{b}(e^{bT} - 1)$$

## The Simplest Model

Although the Gompertz model is the classical approach in expressing mortality, its form is not so convenient for expressing the health state estimates as are presented below. We need a simpler model to express the health status. The best achievement should be to propose a model in which the health measure should be presented by only one main parameter. We thus propose a two parameter model with one crucial health parameter and with similar properties of the Gompertz of the form:

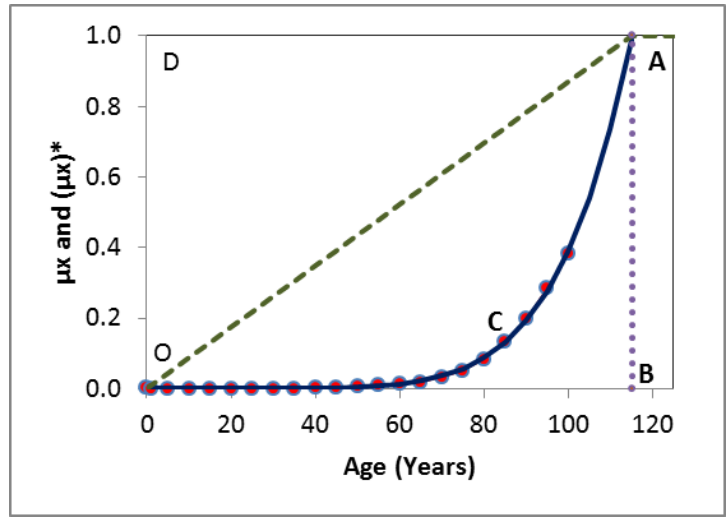

Fig. 1. The mortality diagram

$$\mu_x = \left(\frac{x}{T}\right)^b$$

The parameter $T$ represents the age at which $\mu_x=1$ and $b$ is a crucial health state parameter expressing the curvature of µx. As the health state is improved b gets higher values. The main task is to find the area $E_x$ under the curve OCABO in the mortality diagram (see Figure 1) which is a measure of the mortality effect. This is done by estimating the integral

$$E_x = \int_0^T \left(\frac{x}{T}\right)^b dx = \frac{T}{(b+1)}\left(\frac{x}{T}\right)^b$$

The resulting value for Ex in the interval [0, T] is given by the simple form:

$$E_{mortality} = \frac{T}{(b+1)}$$

It is clear that the total information for the mortality is the area provided under the curve $\mu_x$ and the horizontal axis. The total area $E_{total}$ of the healthy and mortality part of the life span is nothing else but the area included into the rectangle of length T and height 1 that is $E_{total}=T$. The health area is given by

$$E_{health} = T - E_{mortality} = T - \frac{T}{(b+1)} = \frac{bT}{b+1}$$

Then a very simple relation arises for the fraction $E_{health}/E_{mortality}$ that is

$$\frac{E_{health}}{E_{mortality}} = b$$

This is the simplest indicator for the loss of health status of a population. As we have estimated by another method it is more close to the severe disability causes indicator.

The relation $E_{total}/E_{mortality}$ provides another interesting indicator of the form:

$$\frac{E_{total}}{E_{mortality}} = b + 1$$

This indicator is more appropriate for the severe and moderate disability causes indicator (It is compatible with our estimates using the health state approach). It provides larger values for the disability measures as the $E_{total}$ is larger or the $E_{mortality}$ area is smaller by means that as we live longer the disability period becomes larger.

This method suggests a simple but yet interesting tool for classification of various countries and populations, for the loss of healthy life years. A correction multiplier $\lambda$ should be added for specific situations so that the estimator of the loss of healthy life years should be of the form:

$$LHLY = \frac{E_{total}}{E_{mortality}} = \lambda(b+1)$$

However, for comparisons between countries it is sufficient to select $\lambda=1$. Evenmore the selection of $\lambda=1$ is appropriate when we would like to develop a quantitative measure for the LHLY without introducing the public opinion for the health status and the estimates for the cause of diseases and other disability measures. From another point of view the influence of the health status of the society to the public opinions related to health may cause differences in the values for LHLY estimated with the HALE method thus a value for $\lambda$ larger or smaller than unity is needed. By means that we will have to measure not exactly the health status but the public opinion related to the health status, the latter leading in a variety of health estimates in connection to socioeconomic and political situation along with crucial health information from the mass media. Both measures, the standard measure with λ=1 and the flexible one with λ different from 1 could be useful for decision makes and health policy administrators and governmental planners.

To our great surprise our model by selecting λ=1 provided results very close to those provided by WHO as it is presented in the following Tables and in other applications. It is clear that we have found an interesting estimator for the loss of healthy life years.

Our idea to find the loss of healthy life years as a fraction of surfaces in a mortality diagram was proven to be quite important for expressing the health state measures. A more detailed method based on the health state stochastic theory is presented in the book on The Health State Function of a Population and related publications (see Skiadas and Skiadas 2010, 2012, 2015) where more health estimators are found.

**Stability of the coefficients**

The simple model proposed is applied to data by using a non-linear regression analysis technique by using a Levenberg-Marquardt algorithm. The data are obtained from the WHO database providing abridged life tables of the 0-100 years form. The important part of the model is the parameter b expressing the loss of healthy life years. Evenmore *b* can express the curvature of mortality function $\mu_x$. Applying the model to data we need a measure for the selection of the most appropriate value for b.

***When b should be accepted:***

The simpler is to find if b follows a systematic change versus age. We start by selecting all the n data points ($m_0$, $m_1$,…, $m_n$) for $\mu_x$ to find *b* and then we select n-1, n-2,…, n-m for a sufficient number of m<n. As is presented in Figure 2 the parameter *b* follows a systematic change. The example is for USA males and females the year 2000 and the data are from the full life tables of the Human Mortality Database. As it is expected *b* is larger for females than for males. In both cases a distinct maximum value in a specific year of age appears. Accordingly a specific minimum appears for the other not so important parameter *T* (see Figure 3). It is clear that only the specific maximum value for *b* should be selected. Evenmore the estimates for the maximum b account for a local minimum for the first difference dx' of dx provided from the life table. Next Figure 4 illustrates this case for USA males the year 2000 along with a fit curve from our model SK-6. The maximum b is at 94 years for males and females the same as for the minimum of the first difference corresponding to the right inflection point of the death curve dx. Table I includes the parameter estimates for *b* and *T* the year 2000 for USA males and females.

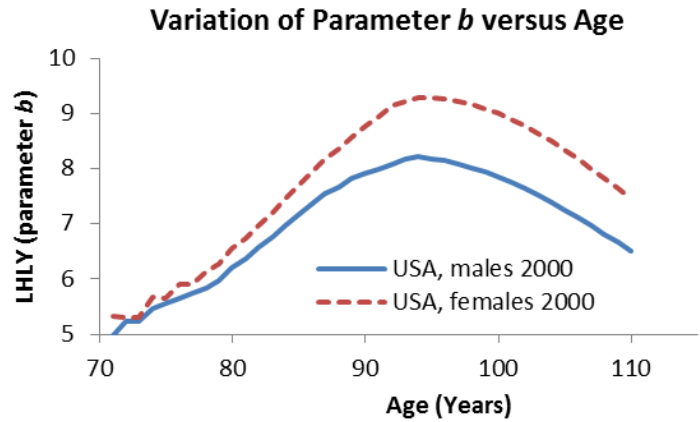

Fig. 2

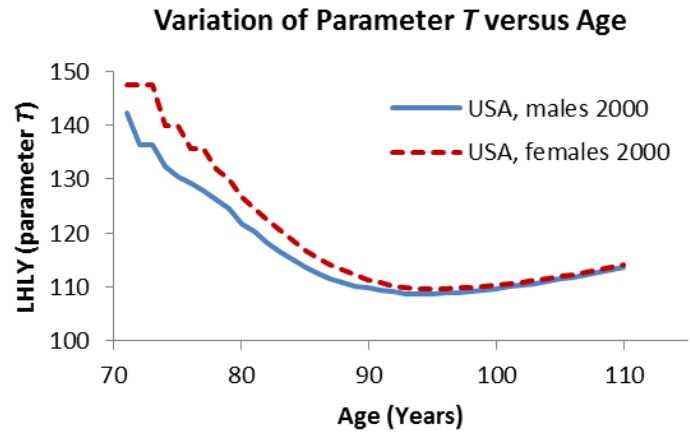

Fig. 3

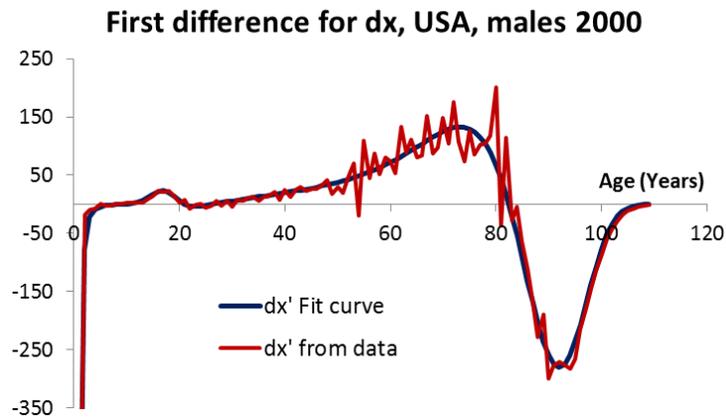

Fig. 4

TABLE I

| Parameter estimates for the model (USA, 2000) ||||||||||
| Age | Females || Males || Age | Females || Males ||
| Years | b | T | b | T | Years | b | T | b | T |
| --- | --- | --- | --- | --- | --- | --- | --- | --- | --- |
| 71 | 5.318 | 147.5 | 4.975 | 142.3 | 91 | 8.942 | 110.7 | 7.992 | 109.4 |
| 72 | 5.308 | 147.5 | 5.244 | 136.4 | 92 | 9.143 | 110.0 | 8.081 | 109.1 |
| 73 | 5.296 | 147.5 | 5.231 | 136.4 | 93 | 9.224 | 109.8 | 8.173 | 108.8 |
| 74 | 5.663 | 140.0 | 5.459 | 132.3 | 94 | **9.291** | **109.6** | **8.218** | **108.6** |
| 75 | 5.649 | 140.0 | 5.559 | 130.5 | 95 | 9.286 | 109.6 | 8.189 | 108.7 |
| 76 | 5.905 | 135.6 | 5.642 | 129.2 | 96 | 9.263 | 109.6 | 8.148 | 108.8 |
| 77 | 5.896 | 135.6 | 5.736 | 127.8 | 97 | 9.224 | 109.7 | 8.094 | 109.0 |
| 78 | 6.146 | 131.9 | 5.844 | 126.3 | 98 | 9.167 | 109.9 | 8.027 | 109.2 |
| 79 | 6.280 | 130.1 | 5.981 | 124.5 | 99 | 9.093 | 110.1 | 7.947 | 109.4 |
| 80 | 6.551 | 126.8 | 6.214 | 121.8 | 100 | 9.002 | 110.3 | 7.856 | 109.7 |
| 81 | 6.748 | 124.6 | 6.368 | 120.2 | 101 | 8.896 | 110.6 | 7.754 | 110.0 |
| 82 | 6.972 | 122.5 | 6.587 | 118.2 | 102 | 8.775 | 110.8 | 7.642 | 110.3 |
| 83 | 7.209 | 120.4 | 6.774 | 116.6 | 103 | 8.641 | 111.2 | 7.521 | 110.7 |
| 84 | 7.453 | 118.5 | 6.981 | 115.0 | 104 | 8.495 | 111.5 | 7.391 | 111.0 |
| 85 | 7.710 | 116.8 | 7.186 | 113.6 | 105 | 8.339 | 111.9 | 7.255 | 111.4 |
| 86 | 7.947 | 115.3 | 7.378 | 112.5 | 106 | 8.173 | 112.3 | 7.114 | 111.8 |
| 87 | 8.185 | 114.0 | 7.546 | 111.5 | 107 | 8.000 | 112.7 | 6.967 | 112.3 |
| 88 | 8.369 | 113.1 | 7.665 | 110.9 | 108 | 7.822 | 113.1 | 6.818 | 112.7 |
| 89 | 8.579 | 112.2 | 7.826 | 110.1 | 109 | 7.638 | 113.5 | 6.666 | 113.2 |
| 90 | 8.778 | 111.3 | 7.916 | 109.8 | 110 | 7.452 | 114.0 | 6.512 | 113.6 |

**The Health State Model (HSM)**

Considering the high importance of the proposed model and the related indicator for the verification of the GBD results we proceed in the introduction of a second method based on the health state of the population instead of the previous one which was based on mortality. This model was proposed earlier (see Skiadas and Skiadas, 2010, 2012, 2013, 2014). This works were based on an earlier publication modeling the health state of a population via a first exit time stochastic methodology. Here we develop a special application adapted to WHO data provided as abridged life tables (0 to 100 with 5 year periods). First we expand the abridged life table to full and then we estimate the health indicators and finally the loss of healthy life year indicators.

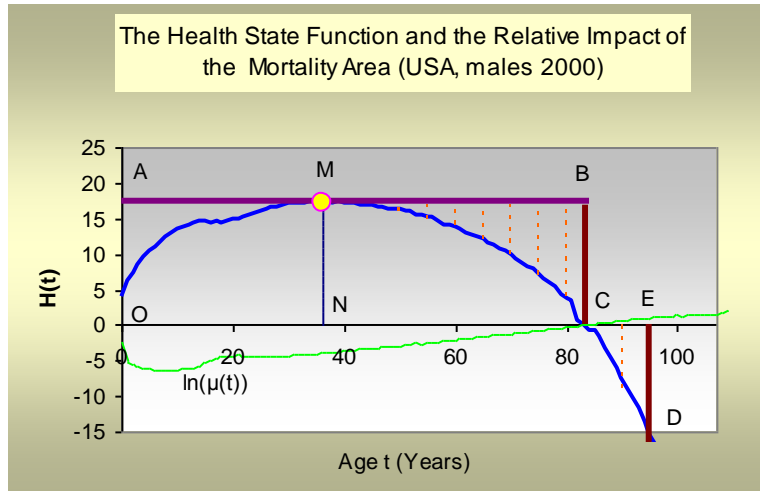

Fig. 5. The impact of the mortality area to health state

By observing the above graph (Figure 5) we can immediately see that the area between the health state curve and the horizontal axis (OMCO) represents the total health dynamics (THD) of the population. Of particular importance is also the area of the health rectangle (OABC) which includes the health state curve. This rectangle is divided in two rectangular parts the smaller (OAMN) indicating the first part of the human life until reaching the point M at the highest level of health state (usually the maximum is between 30 to 45 years) and the second part (NMBC) characterized by the gradual deterioration of the human organism until the zero level of the health state. This zero point health age C is associated with the maximum death rate. After this point the health state level appears as negative in the graph and characterizes a part of the human life totally unstable with high mortality; this is also indicated by a positively increasing form of the logarithm of the force of mortality $\ln(\mu_x)$.

We call the second rectangle NMBC as the **deterioration rectangle**. Instead the first rectangle OAMN is here called as the **development rectangle**. For both cases we can find the relative impact of the area inside each rectangle but outside the health state area to the overall health state. In this study we analyze the relative impact of the **deterioration area** MBCM indicated by dashed lines in the **deterioration rectangle**. It should be noted that if no-deterioration mechanism was present or the repairing mechanism was perfect the health state should continue following the straight line AMB parallel to the X-axis at the level of the maximum health state. The smaller the deterioration area related to the health state area, the higher the healthy life of the population. This comparison can be done by estimating the related areas and making a simple division.

However, when trying to expand the human life further than the limits set by the deterioration mechanisms the percentage of the non-healthy life years becomes higher. This means that we need to divide the total rectangle area by that of the deterioration area to find an estimate for the "lost healthy life years". It is clear that if we don't correct the deterioration mechanisms the loss of healthy years will become higher as the expectation of life becomes larger. This is already observed in the estimates of the World Health Organization (WHO) in the World Health Report for 2000 where the lost healthy years for females are higher than the corresponding values for males. The females show higher life expectancy than males but also higher values for the lost healthy years. The proposed "loss of healthy life years" indicator is given by:

$$LHLY_1 = \lambda \frac{OABC}{THD_{ideal}} \cdot \frac{THD_{ideal}}{MBCM} = \lambda \frac{OABC}{MBCM}$$

Where $THD_{ideal}$ is ideal total health dynamics of the population and the parameter $\lambda$ expresses years and should be estimated according to the specific case. For comparing the related results in various countries we can set $\lambda=1$. When OABC approaches the $THD_{ideal}$ as is the case of several countries in nowadays the loss of healthy life years indicator LHLY can be expressed by other forms.

Another point is the use of the (ECD) area in improving forecasts especially when using the 5-year life tables as is the case of the data for all the WHO Countries. In this case the expanded loss of healthy life years indicator LHLY will take the following two forms:

$$LHLY_2 = \lambda \frac{OMCO + ECD}{MBCM}$$

$$LHLY_3 = \lambda \frac{OABC + ECD}{MBCM}$$

It is clear that the last form will give higher values than the previous one. The following scheme applies: $LHLY_1 < LHLY_2 < LHLY_3$. It remains to explore the forecasting ability of the three forms of the "loss of healthy life years" indicator by applying LHLY to life tables provided by WHO or by the Human Mortality Database or by other sources.

As for the previous case here important is the loss of health state area MBCM whereas the total area including the healthy and non-healthy part is included in OABC+ECD.

$$LHLY_3 = \lambda \frac{OABC + ECD}{MBCM}$$

Details and applications are included in the book on "The Health State Function of a Population", the supplement of this book and other publications (see Skiadas and Skiadas 2010, 2012, 2013, 2016). It is important that we can explore the health state of a population by using the mortality approach with the Simple Model proposed herewith and the health state function approach as well. The latter method provides many important health measures than the simple model.

## TABLE II

| Sex/Region | Healthy Life Expectancy at Birth | | | | | | Life Expectancy at Birth (LE) | | | |
|---|---|---|---|---|---|---|---|---|---|---|
| | 2000 | | | 2012 | | | 2000 | | 2012 | |
| | WHO (HALE) | Mortality Model | HSM Model | WHO (HALE) | Mortality Model | HSM Model | WHO | Mortality Model | WHO | Mortality Model |
| **Both sexes combined** | | | | | | | | | | |
| World | 58.0 | 58.4 | **58.2** | 61.7 | 62.5 | **61.9** | 66.2 | 66.2 | 70.3 | 70.3 |
| High income countries | 67.3 | **67.1** | 67.0 | 69.8 | **69.6** | 69.2 | 76.0 | 76.0 | 78.9 | 78.9 |
| African Region | 43.1 | **42.8** | 42.8 | 49.6 | 49.9 | **49.6** | 50.2 | 50.2 | 57.7 | 57.7 |
| Region of the Americas | 64.9 | 65.7 | **65.4** | 67.1 | 67.7 | **67.2** | 73.9 | 73.9 | 76.4 | 76.3 |
| Eastern Mediterranean Region | 55.4 | 56.9 | **56.6** | 58.3 | 59.7 | **59.4** | 64.9 | 64.9 | 67.8 | 67.8 |
| European Region | 63.9 | 63.9 | 63.9 | 66.9 | 67.2 | **67.0** | 72.4 | 72.4 | 76.1 | 76.0 |
| South East Asian Region | 54.2 | 56.3 | **55.6** | 58.5 | 60.6 | **60.0** | 62.9 | 63.0 | 67.5 | 67.5 |
| Western Pacific Region | 64.8 | **63.9** | **64.2** | 68.1 | **67.3** | 67.5 | 72.3 | 72.3 | 75.9 | 75.9 |
| **Males** | | | | | | | | | | |
| World | 56.4 | **56.6** | 56.2 | 60.1 | 60.4 | **60.0** | 63.9 | 63.9 | 68.1 | 68.0 |
| High income countries | 64.7 | 64.1 | **64.2** | 67.5 | **67.0** | 67.0 | 72.4 | 72.3 | 75.8 | 75.7 |
| African Region | 42.4 | 41.6 | **42.3** | 48.8 | **48.6** | **48.6** | 49.0 | 49.0 | 56.3 | 56.3 |
| Region of the Americas | 62.7 | 63.1 | **62.5** | 64.9 | 65.1 | **64.6** | 70.8 | 70.8 | 73.5 | 73.5 |
| Eastern Mediterranean Region | 54.8 | 55.7 | **55.6** | 57.4 | 58.2 | **57.9** | 63.6 | 63.6 | 66.1 | 66.1 |
| European Region | 60.7 | **60.4** | 61.1 | 64.2 | **64.3** | 64.5 | 68.2 | 68.2 | 72.4 | 72.4 |
| South East Asian Region | 53.5 | 55.4 | **54.6** | 57.4 | 59.2 | **58.6** | 61.6 | 61.7 | 65.7 | 65.7 |
| Western Pacific Region | 63.0 | 61.8 | **62.0** | 66.6 | 65.2 | **65.7** | 70.0 | 70.0 | 73.9 | 73.9 |
| **Females** | | | | | | | | | | |
| World | 59.7 | 60.3 | **59.9** | 63.4 | 64.3 | **64.1** | 68.5 | 68.5 | 72.7 | 72.6 |
| High income countries | 70.0 | **69.7** | 69.6 | 72.0 | 71.8 | **72.1** | 79.6 | 79.5 | 82.0 | 81.9 |
| African Region | 43.8 | **43.8** | 43.5 | 50.4 | 51.2 | **50.5** | 51.4 | 51.4 | 59.0 | 59.1 |
| Region of the Americas | 67.2 | 68.0 | **67.8** | 69.1 | 69.9 | **69.8** | 77.0 | 76.9 | 79.3 | 79.2 |
| Eastern Mediterranean Region | 56.1 | 58.2 | **57.8** | 59.2 | 61.3 | **61.0** | 66.4 | 66.4 | 69.7 | 69.6 |
| European Region | 67.1 | 67.6 | **67.3** | 69.6 | 70.0 | **69.7** | 76.7 | 76.6 | 79.6 | 79.6 |
| South East Asian Region | 55.0 | 57.2 | **56.4** | 59.7 | 62.0 | **61.7** | 64.3 | 64.4 | 69.4 | 69.4 |
| Western Pacific Region | 66.7 | 65.7 | **66.1** | 69.8 | 68.9 | **69.1** | 74.8 | 74.8 | 78.1 | 78.0 |

**First Application**

The Table II includes our estimates for the healthy life expectancy at birth for the years 2000 and 2012 by applying the proposed mortality model and the health state model (HSM), and the estimates of WHO referred as HALE and included in the WHO websites (August 2015). Our estimates for the mortality model are based on LHLY=(b+1)=Etotal/Emortality.

The main finding is that our models verify the WHO (HALE) estimates based on the Global Burden of Disease Study. Our results are quite close (with less to one year difference) to the estimates for the World, the High Income Countries, the African region, the European region and Western Pacific and differ by 1-2 years for the Eastern Mediterranean region and the South East Asian region. In the last two cases the collection of data and the accuracy of the information sources may lead to high uncertainty of the related health state estimates. This is demonstrated in the provided confidence intervals for the estimates in countries of these regions in the studies by Salomon et al. (2012) and the Report of WHO (2001) for the HLE of the member states (2000). From the Salomon et al. study we have calculated a mean confidence interval of 5.5 years for males and 6.8 for females for the year 2000. We thus propose to base the future works on the system we propose and to use it to calibrate the estimates especially for the countries providing of low accuracy data.

To support future studies we have formulated an easy to use framework in Excel. The only needed is to insert data for μx in the related column of the program. The program estimates the life expectancy, the loss of healthy life years and the healthy life expectancy.

The program includes the Figure 1 providing the fit curve (solid line) to the provided data for μx (dotted line) while the straight dashed line is the limit for μx representing a simple decay process. In the latter case the parameter b=1.

**Second Application**

Another application is presented in Table III where the mortality model and the WHO (HALE) results from 1990 to 2013 are compared for the WHO member countries.

TABLE III

Mean values of the Mortality Model for the Loss of Healthy Life Years (LHLY) and the related results from the HALE method for the WHO countries

| Type | HALE | MODEL | HALEb | HALE | MODEL | HALE | HALE | HALE | HALE | MODEL | HALE | MODEL |
|---|---|---|---|---|---|---|---|---|---|---|---|---|
| Year | 1990 | 1990 | 2000 | 2000 | 2000 | 2001 | 2002 | 2010 | 2012 | 2012 | 2013 | 2013 |
| Males | 8.7 | 7.7 | 8.4 | 8.2 | 7.8 | 9.7 | 7.2 | 9.3 | 8.8 | 8.2 | 8.9 | 8.3 |
| Females | 10.3 | 8.6 | 11.0 | 9.5 | 8.8 | 10.0 | 8.9 | 10.8 | 10.2 | 9.3 | 10.1 | 9.3 |

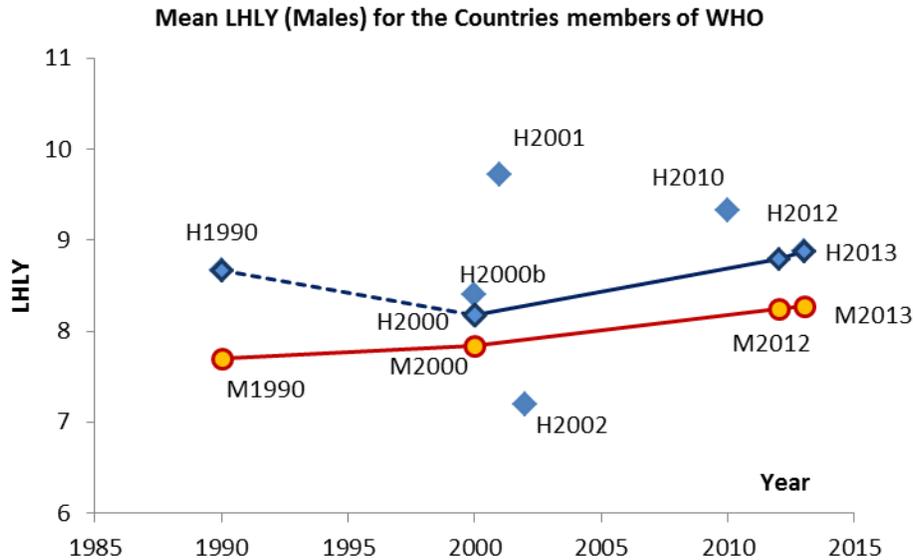

Fig.6

Figure 6 and Figure 7 illustrate the estimates of the arithmetic mean the loss of healthy life years (LHLY) for various time periods with the proposed mortality model (circles) versus the HALE method (rhombus) for the countries of WHO (males). The model estimates and the HALE differ by 1.0 years in 1990, 0.4 years in 2000, 0.6 years in 2012 and 0.6 years in 2013. Several options of the related figures for HALE are included starting from the first estimates in 190 until the estimates of 2013. The estimates for 1990, 2000, 2012 and 2013 are provided in the last WHO websites and can be accepted as the official estimates whereas the HALE estimates for 2001, 2002 and 2010 are also included. It should be noted that significant differences appear in HALE estimates in the group of cases of 2000, 2001 and 2002 and 2010 and 2012 due to improvements in the methodology and the use of new epidemiological data. In the Annex Table of the World Health Report 2001 and the related of 2002 write:

*Healthy life expectancy estimates published here are not directly comparable to those published in the World Health Report 2000, due to improvements in survey methodology and the use of new epidemiological data for some diseases. See Statistical Annex notes (pp.130–135). The figures reported in this Table along with the data collection and estimation methods have been largely developed by WHO and do not necessarily reflect official statistics of Member States. Further development in collaboration with Member States is underway for improved data collection and estimation methods (WHO 2001).*
*Healthy life expectancy estimates published here are not directly comparable to those published in The World Health Report 2001, because of improvements in survey methodology and the use of new epidemiological data for some diseases and revisions of life tables for 2000 for many Member States to take new data into account (see Statistical Annex explanatory notes). The figures reported in this Table along with the data collection and estimation methods have been largely developed by WHO and do not necessarily reflect official statistics of Member States. Further development in collaboration with Member States is under way for improved data collection and estimation methods (WHO 2002).*

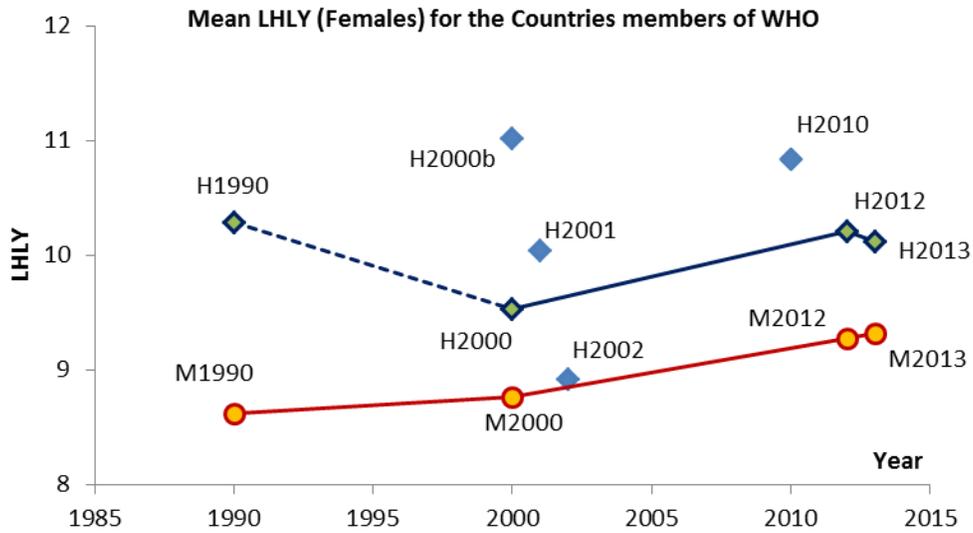

Fig. 7

Figure 7 summarizes the estimates of the mean of the loss of healthy life years (LHLY) for various time periods with the proposed model (circles) versus the HALE method (rhombus) for the countries of WHO (females). The model estimates (females) and the HALE differ by 1.7 years in 1990, 0.7 years in 2000, 0.9 years in 2012 and 0.8 years in 2013. For females as for males the estimated differences between model and HALE are higher in 1990 than for the following years due to the higher values of the HALE estimates.

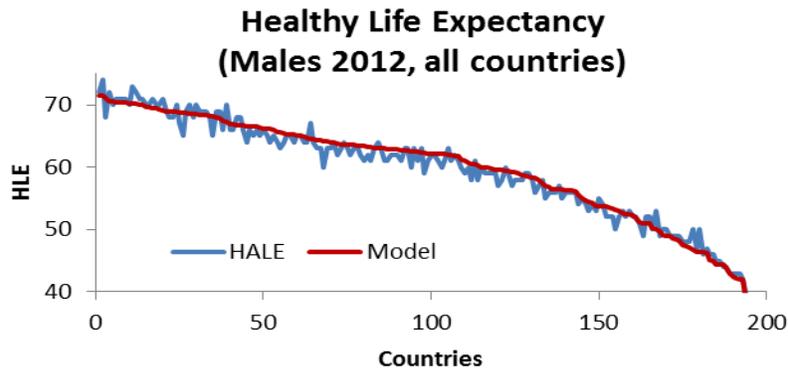

Fig. 8

Figure 8 illustrates the Healthy life expectancy (HLE) for males (2012) for all the WHO countries, estimated by the Model (red line) and HALE estimates (Blue line). The mean value is 59.6 for HALE and 60.2 years for the model.

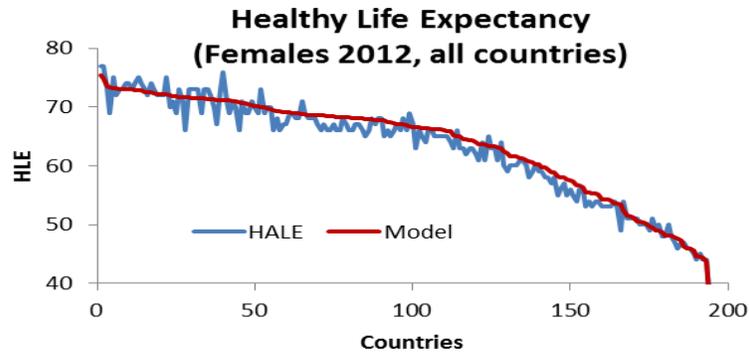

Fig. 9

Figure 9 illustrates the Healthy life expectancy (HLE) for females (2012) for all the WHO countries, estimated by the Model (red line) and HALE estimates (Blue line). The mean value is 63.1 for HALE and 63.8 years for the model.

The full estimated figures are included in Table IV, Table V and Table VI in the end of the paper.

**Discussion and Conclusions**

The GBD study critisized by Williams (see Murray et al. 2000) whereas many comments from people from social sciences and philosophy refer to the impossibility to define health and, as a consequence, to measure it. The main problem is that we cannot have flexibility in finding an estimate of health the way we do with other measures of the human organism and related activities. So far if we measure health by collecting surveys it is clear that the uncertainty is relatively high. Even more if we decide for an accepted health state estimate (see Sanders, 1964 and related studies during 60's and 70') it remains the problem of accepting a *unit of measure*. The method we propose overcomes many of the objections posed while is simple and easy to apply.

# TABLE IV

| Country - 2012 - Females | HALE | MODEL | Country - 2012 - Males | HALE | MODEL |
|---|---|---|---|---|---|
| Afghanistan | 49 | 53.4 | Afghanistan | 49 | 51.0 |
| Albania | 66 | 65.1 | Albania | 64 | 63.9 |
| Algeria | 63 | 64.4 | Algeria | 62 | 61.9 |
| Andorra | 74 | 73.0 | Andorra | 70 | 68.6 |
| Angola | 45 | 44.5 | Angola | 43 | 42.2 |
| Antigua and Barbuda | 66 | 67.4 | Antigua and Barbuda | 63 | 64.0 |
| Argentina | 69 | 69.9 | Argentina | 64 | 65.2 |
| Armenia | 66 | 66.2 | Armenia | 60 | 59.9 |
| Australia | 74 | 72.8 | Australia | 71 | 70.4 |
| Austria | 73 | 71.5 | Austria | 69 | 68.3 |
| Azerbaijan | 65 | 66.4 | Azerbaijan | 61 | 62.2 |
| Bahamas | 67 | 67.1 | Bahamas | 62 | 62.9 |
| Bahrain | 66 | 68.5 | Bahrain | 66 | 66.9 |
| Bangladesh | 61 | 63.7 | Bangladesh | 60 | 62.5 |
| Barbados | 69 | 70.5 | Barbados | 64 | 66.2 |
| Belarus | 68 | 68.6 | Belarus | 59 | 59.9 |
| Belgium | 73 | 71.4 | Belgium | 69 | 67.9 |
| Belize | 66 | 68.1 | Belize | 61 | 62.2 |
| Benin | 51 | 51.4 | Benin | 50 | 49.1 |
| Bhutan | 59 | 62.1 | Bhutan | 58 | 60.3 |
| Bolivia (Plurinational State of) | 61 | 61.2 | Bolivia (Plurinational State of) | 58 | 57.5 |
| Bosnia and Herzegovina | 70 | 70.2 | Bosnia and Herzegovina | 66 | 66.7 |
| Botswana | 53 | 55.3 | Botswana | 52 | 53.4 |
| Brazil | 67 | 69.2 | Brazil | 62 | 63.3 |
| Brunei Darussalam | 69 | 69.0 | Brunei Darussalam | 68 | 66.8 |
| Bulgaria | 68 | 67.7 | Bulgaria | 63 | 63.0 |
| Burkina Faso | 51 | 50.5 | Burkina Faso | 50 | 49.0 |
| Burundi | 49 | 49.7 | Burundi | 46 | 46.4 |
| Cabo Verde | 66 | 68.2 | Cabo Verde | 61 | 62.4 |
| Cambodia | 63 | 66.5 | Cambodia | 59 | 62.3 |
| Cameroon | 49 | 49.1 | Cameroon | 48 | 47.3 |
| Canada | 73 | 73.0 | Canada | 71 | 70.0 |
| Central African Republic | 44 | 43.7 | Central African Republic | 43 | 42.1 |
| Chad | 44 | 44.4 | Chad | 43 | 42.9 |
| Chile | 72 | 73.2 | Chile | 68 | 68.9 |
| China | 69 | 66.7 | China | 67 | 64.6 |
| Colombia | 70 | 72.0 | Colombia | 66 | 66.6 |
| Comoros | 54 | 55.2 | Comoros | 53 | 52.7 |
| Congo | 51 | 51.5 | Congo | 49 | 49.0 |
| Cook Islands | 66 | 67.9 | Cook Islands | 63 | 65.6 |
| Costa Rica | 71 | 71.9 | Costa Rica | 68 | 68.9 |
| Côte d'Ivoire | 46 | 45.9 | Côte d'Ivoire | 45 | 44.4 |
| Croatia | 70 | 71.0 | Croatia | 65 | 66.5 |
| Cuba | 69 | 71.9 | Cuba | 65 | 68.7 |
| Cyprus | 76 | 71.1 | Cyprus | 73 | 70.2 |
| Czech Republic | 71 | 70.4 | Czech Republic | 66 | 66.2 |
| Democratic People's Republic of Korea | 65 | 63.4 | Democratic People's Republic of Korea | 59 | 58.4 |
| Democratic Republic of the Congo | 45 | 45.6 | Democratic Republic of the Congo | 43 | 42.4 |
| Denmark | 72 | 72.2 | Denmark | 69 | 68.9 |
| Djibouti | 53 | 54.3 | Djibouti | 52 | 51.8 |
| Dominica | 65 | 68.0 | Dominica | 61 | 63.2 |
| Dominican Republic | 67 | 69.2 | Dominican Republic | 65 | 68.1 |
| Ecuador | 68 | 67.7 | Ecuador | 64 | 63.1 |
| Egypt | 63 | 65.2 | Egypt | 60 | 61.2 |
| El Salvador | 66 | 69.3 | El Salvador | 59 | 61.0 |
| Equatorial Guinea | 48 | 48.7 | Equatorial Guinea | 47 | 46.2 |
| Eritrea | 55 | 58.1 | Eritrea | 53 | 54.4 |
| Estonia | 71 | 71.0 | Estonia | 63 | 63.5 |
| Ethiopia | 56 | 56.2 | Ethiopia | 54 | 53.7 |
| Fiji | 62 | 64.4 | Fiji | 58 | 59.6 |
| Finland | 73 | 71.6 | Finland | 69 | 67.7 |

| Country | | | Country | | |
| --- | --- | --- | --- | --- | --- |
| France | 74 | 72.5 | France | 69 | 68.3 |
| Gabon | 54 | 55.1 | Gabon | 53 | 53.8 |
| Gambia | 53 | 53.7 | Gambia | 52 | 50.9 |
| Georgia | 68 | 68.6 | Georgia | 62 | 62.8 |
| Germany | 73 | 71.5 | Germany | 70 | 68.5 |
| Ghana | 54 | 52.5 | Ghana | 53 | 50.1 |
| Greece | 73 | 71.1 | Greece | 69 | 68.7 |
| Grenada | 66 | 67.3 | Grenada | 60 | 62.1 |
| Guatemala | 65 | 65.9 | Guatemala | 60 | 59.5 |
| Guinea | 50 | 50.3 | Guinea | 49 | 48.6 |
| Guinea-Bissau | 47 | 47.4 | Guinea-Bissau | 46 | 45.1 |
| Guyana | 57 | 58.9 | Guyana | 52 | 52.5 |
| Haiti | 53 | 55.5 | Haiti | 50 | 53.1 |
| Honduras | 65 | 67.4 | Honduras | 62 | 63.6 |
| Hungary | 69 | 69.1 | Hungary | 63 | 64.2 |
| Iceland | 73 | 71.5 | Iceland | 72 | 70.2 |
| India | 58 | 60.8 | India | 56 | 58.2 |
| Indonesia | 64 | 63.7 | Indonesia | 61 | 60.4 |
| Iran (Islamic Republic of) | 65 | 66.1 | Iran (Islamic Republic of) | 63 | 63.7 |
| Iraq | 63 | 64.4 | Iraq | 58 | 58.4 |
| Ireland | 73 | 73.0 | Ireland | 70 | 69.7 |
| Israel | 73 | 72.5 | Israel | 71 | 70.3 |
| Italy | 74 | 73.0 | Italy | 71 | 70.3 |
| Jamaica | 66 | 66.9 | Jamaica | 62 | 62.2 |
| Japan | 77 | 75.0 | Japan | 72 | 70.5 |
| Jordan | 65 | 66.2 | Jordan | 64 | 63.6 |
| Kazakhstan | 64 | 63.0 | Kazakhstan | 56 | 56.5 |
| Kenya | 54 | 53.7 | Kenya | 52 | 51.0 |
| Kiribati | 60 | 60.3 | Kiribati | 56 | 56.4 |
| Kuwait | 67 | 68.5 | Kuwait | 68 | 68.5 |
| Kyrgyzstan | 63 | 64.6 | Kyrgyzstan | 58 | 58.7 |
| Lao People's Democratic Republic | 58 | 58.9 | Lao People's Democratic Republic | 56 | 56.5 |
| Latvia | 68 | 68.9 | Latvia | 61 | 62.0 |
| Lebanon | 71 | 71.7 | Lebanon | 68 | 68.3 |
| Lesotho | 44 | 44.5 | Lesotho | 42 | 42.0 |
| Liberia | 53 | 54.1 | Liberia | 52 | 52.2 |
| Libya | 65 | 67.7 | Libya | 64 | 64.5 |
| Lithuania | 70 | 69.6 | Lithuania | 61 | 61.8 |
| Luxembourg | 73 | 73.5 | Luxembourg | 70 | 70.5 |
| Madagascar | 56 | 58.0 | Madagascar | 54 | 55.5 |
| Malawi | 51 | 49.6 | Malawi | 50 | 46.8 |
| Malaysia | 66 | 67.1 | Malaysia | 63 | 64.3 |
| Maldives | 67 | 68.1 | Maldives | 66 | 67.0 |
| Mali | 48 | 48.1 | Mali | 49 | 48.4 |
| Malta | 72 | 71.4 | Malta | 70 | 69.3 |
| Marshall Islands | 61 | 63.2 | Marshall Islands | 57 | 59.6 |
| Mauritania | 54 | 55.5 | Mauritania | 52 | 52.9 |
| Mauritius | 68 | 68.3 | Mauritius | 62 | 62.2 |
| Mexico | 69 | 70.2 | Mexico | 65 | 65.1 |
| Micronesia (Federated States of) | 60 | 61.6 | Micronesia (Federated States of) | 59 | 59.7 |
| Monaco | 75 | 72.1 | Monaco | 70 | 67.2 |
| Mongolia | 63 | 63.3 | Mongolia | 56 | 56.2 |
| Montenegro | 67 | 68.6 | Montenegro | 65 | 65.3 |
| Morocco | 61 | 63.6 | Morocco | 60 | 60.7 |
| Mozambique | 46 | 45.9 | Mozambique | 45 | 44.4 |
| Myanmar | 58 | 59.3 | Myanmar | 56 | 56.1 |
| Namibia | 59 | 60.5 | Namibia | 55 | 56.4 |
| Nauru | 69 | 73.3 | Nauru | 64 | 66.6 |
| Nepal | 60 | 60.9 | Nepal | 58 | 59.0 |
| Netherlands | 72 | 72.2 | Netherlands | 70 | 70.3 |
| New Zealand | 73 | 71.4 | New Zealand | 71 | 69.1 |
| Nicaragua | 66 | 68.6 | Nicaragua | 61 | 63.0 |
| Niger | 50 | 50.4 | Niger | 50 | 50.2 |
| Nigeria | 47 | 46.3 | Nigeria | 46 | 44.9 |
| Niue | 66 | 68.3 | Niue | 62 | 63.4 |

| Country | | | Country | | |
|---|---|---|---|---|---|
| Norway | 72 | 72.2 | Norway | 70 | 69.4 |
| Oman | 67 | 71.2 | Oman | 65 | 66.4 |
| Pakistan | 57 | 57.9 | Pakistan | 56 | 56.4 |
| Palau | 64 | 65.8 | Palau | 61 | 62.7 |
| Panama | 69 | 71.1 | Panama | 65 | 66.0 |
| Papua New Guinea | 55 | 57.7 | Papua New Guinea | 52 | 53.6 |
| Paraguay | 67 | 68.0 | Paraguay | 63 | 63.1 |
| Peru | 68 | 69.4 | Peru | 66 | 66.5 |
| Philippines | 63 | 63.4 | Philippines | 57 | 57.9 |
| Poland | 71 | 71.1 | Poland | 64 | 65.5 |
| Portugal | 73 | 73.0 | Portugal | 69 | 68.4 |
| Qatar | 66 | 71.6 | Qatar | 68 | 71.1 |
| Republic of Korea | 75 | 72.8 | Republic of Korea | 70 | 68.8 |
| Republic of Moldova | 66 | 66.5 | Republic of Moldova | 59 | 59.9 |
| Romania | 69 | 68.8 | Romania | 63 | 62.6 |
| Russian Federation | 66 | 66.3 | Russian Federation | 57 | 56.4 |
| Rwanda | 56 | 57.6 | Rwanda | 55 | 55.1 |
| Saint Kitts and Nevis | 66 | 68.0 | Saint Kitts and Nevis | 61 | 62.9 |
| Saint Lucia | 66 | 70.8 | Saint Lucia | 60 | 64.2 |
| Saint Vincent and the Grenadines | 65 | 66.1 | Saint Vincent and the Grenadines | 61 | 63.4 |
| Samoa | 66 | 68.4 | Samoa | 62 | 62.9 |
| San Marino | 73 | 69.9 | San Marino | 72 | 71.5 |
| Sao Tome and Principe | 59 | 60.2 | Sao Tome and Principe | 56 | 56.8 |
| Saudi Arabia | 66 | 68.3 | Saudi Arabia | 64 | 64.8 |
| Senegal | 56 | 56.6 | Senegal | 54 | 54.3 |
| Serbia | 67 | 68.2 | Serbia | 63 | 64.3 |
| Seychelles | 71 | 68.8 | Seychelles | 63 | 62.0 |
| Sierra Leone | 39 | 38.2 | Sierra Leone | 39 | 38.1 |
| Singapore | 77 | 75.3 | Singapore | 74 | 71.4 |
| Slovakia | 70 | 69.8 | Slovakia | 64 | 64.9 |
| Slovenia | 73 | 72.3 | Slovenia | 67 | 68.7 |
| Solomon Islands | 60 | 61.6 | Solomon Islands | 58 | 59.2 |
| Somalia | 46 | 47.6 | Somalia | 44 | 43.8 |
| South Africa | 53 | 54.3 | South Africa | 49 | 49.6 |
| South Sudan | 48 | 48.2 | South Sudan | 47 | 46.4 |
| Spain | 75 | 73.3 | Spain | 71 | 70.0 |
| Sri Lanka | 68 | 68.6 | Sri Lanka | 63 | 63.5 |
| Sudan | 54 | 56.7 | Sudan | 52 | 53.6 |
| Suriname | 68 | 67.1 | Suriname | 63 | 63.6 |
| Swaziland | 47 | 47.8 | Swaziland | 44 | 44.3 |
| Sweden | 73 | 71.7 | Sweden | 71 | 69.5 |
| Switzerland | 74 | 72.8 | Switzerland | 71 | 70.3 |
| Syrian Arab Republic | 65 | 66.1 | Syrian Arab Republic | 55 | 53.7 |
| Tajikistan | 60 | 61.7 | Tajikistan | 59 | 59.6 |
| Thailand | 68 | 69.0 | Thailand | 63 | 62.5 |
| The former Yugoslav Republic of Macedonia | 68 | 67.9 | The former Yugoslav Republic of Macedonia | 65 | 65.3 |
| Timor-Leste | 58 | 59.7 | Timor-Leste | 55 | 57.0 |
| Togo | 50 | 48.1 | Togo | 50 | 46.4 |
| Tonga | 61 | 61.3 | Tonga | 64 | 65.7 |
| Trinidad and Tobago | 64 | 66.3 | Trinidad and Tobago | 58 | 60.4 |
| Tunisia | 67 | 68.4 | Tunisia | 65 | 65.1 |
| Turkey | 67 | 67.8 | Turkey | 63 | 62.4 |
| Turkmenistan | 59 | 59.8 | Turkmenistan | 53 | 52.4 |
| Tuvalu | 60 | 62.4 | Tuvalu | 57 | 59.3 |
| Uganda | 50 | 50.4 | Uganda | 49 | 48.2 |
| Ukraine | 67 | 66.6 | Ukraine | 59 | 58.6 |
| United Arab Emirates | 66 | 69.4 | United Arab Emirates | 66 | 67.4 |
| United Kingdom | 72 | 72.5 | United Kingdom | 70 | 69.6 |
| United Republic of Tanzania | 53 | 54.3 | United Republic of Tanzania | 51 | 51.1 |
| United States of America | 71 | 70.7 | United States of America | 68 | 66.8 |
| Uruguay | 70 | 71.4 | Uruguay | 65 | 66.2 |
| Uzbekistan | 62 | 64.0 | Uzbekistan | 59 | 59.4 |
| Vanuatu | 63 | 65.0 | Vanuatu | 61 | 62.1 |
| Venezuela (Bolivarian Republic of) | 69 | 70.6 | Venezuela (Bolivarian Republic of) | 63 | 62.7 |
| Viet Nam | 69 | 71.5 | Viet Nam | 62 | 63.8 |
| Yemen | 55 | 57.4 | Yemen | 54 | 54.7 |
| Zambia | 50 | 49.0 | Zambia | 48 | 47.5 |
| Zimbabwe | 51 | 51.3 | Zimbabwe | 48 | 47.2 |

TABLE V

Estimates of the Mortality Model for the Loss of Healthy Life Years (LHLY) for males for the WHO member countries and the related results from the HALE method of the World Health Organization

| Countries / Year | HALE 1990 | Model 1990 | HALE 2000 | HALE 2000b | Model 2000 | HALE 2001 | HALE 2002 | HALE 2010 | HALE 2012 | Model 2012 | HALE 2013 | Model 2013 |
|---|---|---|---|---|---|---|---|---|---|---|---|---|
| Afghanistan | 8.9 | 6.9 | 9.1 | 9 | 7.2 | 10.0 | 6.6 | 9.7 | 9 | 7.5 | 11 | 7.5 |
| Albania | 9.3 | 8.6 | 7.9 | 8 | 8.6 | 10.4 | 7.8 | 9.5 | 9 | 8.6 | 9 | 8.6 |
| Algeria | 9.8 | 8.3 | 9.7 | 9 | 8.3 | 11.9 | 7.9 | 10.5 | 8 | 8.3 | 8 | 8.3 |
| Andorra | 10.6 | 8.8 | 7.3 | 9 | 8.7 | 7.4 | 7.0 | 11.5 | 10 | 10.6 | 9 | 10.6 |
| Angola | 6.4 | 7.2 | 8.1 | 6 | 7.4 | 8.4 | 6.3 | 8.2 | 7 | 7.7 | 7 | 7.7 |
| Antigua and Barbuda | 9.7 | 7.6 | 10.1 | 9 | 8.5 | 11.8 | 8.9 | 12.9 | 10 | 9.3 | 10 | 9.2 |
| Argentina | 8.5 | 7.1 | 8.4 | 9 | 7 | 9.5 | 8.3 | 9.0 | 9 | 7.5 | 9 | 7.5 |
| Armenia | 8.7 | 6.6 | 7.5 | 8 | 7.4 | 10.8 | 7.6 | 9.0 | 8 | 7.2 | 8 | 7.3 |
| Australia | 9.7 | 8.1 | 6.9 | 9 | 9.3 | 7.3 | 7.0 | 10.8 | 10 | 10.0 | 9 | 9.9 |
| Austria | 9.1 | 8.4 | 6.8 | 9 | 9.1 | 7.0 | 7.1 | 10.7 | 10 | 9.9 | 11 | 10.0 |
| Azerbaijan | 8.1 | 6.9 | 8.4 | 8 | 6.9 | 10.4 | 7.2 | 9.0 | 8 | 7.3 | 9 | 7.2 |
| Bahamas | 10.8 | 8.6 | 10.8 | 8 | 9.5 | 14.1 | 8.4 | 12.1 | 10 | 9.4 | 11 | 9.4 |
| Bahrain | 9.4 | 7.9 | 9.7 | 10 | 8 | 9.9 | 7.9 | 10.1 | 10 | 9.3 | 10 | 9.3 |
| Bangladesh | 9.4 | 7.4 | 9.8 | 10 | 7.2 | 10.2 | 7.3 | 12.4 | 9 | 6.9 | 10 | 6.9 |
| Barbados | 8.1 | 8.7 | 9.3 | 9 | 8.7 | 9.5 | 7.6 | 7.7 | 11 | 8.8 | 11 | 8.8 |
| Belarus | 9.2 | 6.9 | 6.6 | 8 | 6.5 | 9.0 | 6.1 | 10.2 | 9 | 6.7 | 9 | 6.6 |
| Belgium | 9.9 | 7.9 | 6.9 | 9 | 8.8 | 7.1 | 6.3 | 11.6 | 10 | 9.8 | 9 | 9.8 |
| Belize | 8.2 | 8.7 | 11.1 | 10 | 8.3 | 11.4 | 9.0 | 8.5 | 11 | 9.5 | 11 | 9.5 |
| Benin | 8.6 | 8.0 | 8.5 | 8 | 8 | 10.9 | 6.6 | 9.4 | 7 | 8.3 | 7 | 8.3 |
| Bhutan | 8.7 | 6.5 | 10.3 | 8 | 6.8 | 10.5 | 7.3 | 9.6 | 9 | 7.3 | 9 | 7.3 |
| Bolivia | 9.1 | 7.9 | 9.5 | 8 | 8.1 | 13.1 | 8.2 | 9.7 | 8 | 7.9 | 8 | 8.0 |
| Bosnia and Herzegovina | 9.3 | 7.1 | 6.6 | 9 | 6.2 | 9.3 | 7.0 | 11.0 | 10 | 7.8 | 9 | 7.8 |
| Botswana | 8.8 | 9.7 | 6.5 | 6 | 6.7 | 6.4 | 4.2 | 9.4 | 9 | 7.7 | 10 | 8.0 |
| Brazil | 9.1 | 6.6 | 9.5 | 8 | 6.2 | 13.3 | 8.5 | 9.3 | 8 | 6.9 | 9 | 6.3 |
| Brunei Darussalam | 8.1 | 8.6 | 9.6 | 7 | 8.4 | 12.8 | 9.7 | 8.6 | 8 | 8.8 | 8 | 8.9 |
| Bulgaria | 7.6 | 7.5 | 6.3 | 8 | 7.8 | 7.5 | 6.2 | 7.4 | 9 | 7.7 | 9 | 7.6 |
| Burkina Faso | 7.3 | 7.9 | 7.2 | 7 | 8.1 | 8.3 | 5.6 | 7.5 | 7 | 8.2 | 8 | 8.2 |
| Burundi | 6.6 | 7.3 | 6.7 | 6 | 7.2 | 6.8 | 5.3 | 7.5 | 8 | 7.5 | 7 | 7.6 |
| Cambodia | 8.5 | 6.2 | 7.8 | 9 | 6.4 | 10.3 | 6.3 | 8.7 | 10 | 7.1 | 10 | 7.1 |
| Cameroon | 8.4 | 8.0 | 8.1 | 7 | 7.8 | 10.1 | 6.0 | 8.1 | 11 | 8.0 | 10 | 8.0 |
| Canada | 9.3 | 7.9 | 7.7 | 9 | 8.8 | 8.4 | 7.1 | 10.2 | 7 | 9.5 | 8 | 9.6 |
| Cape Verde | 9.2 | 7.7 | 9.6 | 9 | 7.9 | 13.6 | 7.9 | 10.1 | 10 | 8.3 | 9 | 8.4 |
| Central African Republic | 6.6 | 7.5 | 6.9 | 6 | 7.4 | 9.7 | 5.1 | 5.9 | 7 | 7.8 | 7 | 7.8 |
| Chad | 7.9 | 7.2 | 8.7 | 6 | 7.3 | 11.1 | 6.4 | 8.2 | 7 | 7.5 | 7 | 7.5 |
| Chile | 8.7 | 7.2 | 9 | 9 | 7.2 | 8.7 | 8.5 | 9.3 | 9 | 7.8 | 9 | 7.9 |
| China | 7.2 | 8.0 | 8 | 8 | 8.8 | 7.7 | 6.5 | 7.4 | 7 | 9.3 | 7 | 9.2 |
| Colombia | 8.9 | 8.4 | 8.6 | 9 | 7.1 | 11.4 | 9.7 | 9.3 | 11 | 7.8 | 10 | 9.8 |
| Comoros | 7.4 | 7.2 | 9.1 | 7 | 7.3 | 12.8 | 7.8 | 8.2 | 7 | 7.5 | 7 | 7.5 |
| Congo | 7.6 | 7.9 | 7.7 | 7 | 7.9 | 10.9 | 6.3 | 7.9 | 8 | 8.2 | 8 | 8.2 |
| Cook Islands |  | 7.8 | 8.3 | 10 | 9.3 | 11.6 | 8.6 |  | 10 | 7.7 | 11 | 7.7 |
| Costa Rica | 9.9 | 7.1 | 9.2 | 10 | 7.4 | 11.1 | 9.5 | 9.8 | 9 | 8.0 | 9 | 8.1 |
| Côte d'Ivoire | 7.9 | 7.7 | 7.2 | 6 | 7.5 | 8.7 | 5.4 | 7.4 | 7 | 7.9 | 7 | 7.9 |
| Croatia | 8.8 | 7.0 | 9 | 9 | 7.3 | 9.2 | 7.2 | 9.8 | 9 | 7.5 | 10 | 7.5 |
| Cuba | 9.5 | 8.6 | 8.6 | 10 | 7.7 | 10.0 | 7.9 | 12.6 | 11 | 7.6 | 12 | 8.1 |
| Cyprus | 9.9 | 8.9 | 8.4 | 7 | 10.5 | 9.4 | 8.8 | 10.5 | 8 | 9.4 | 7 | 9.4 |
| Czech Republic | 8.1 | 7.3 | 8.6 | 9 | 8 | 8.1 | 6.6 | 9.5 | 9 | 8.7 | 9 | 8.7 |

| Country | | | | | | | | | | | | |
|---|---|---|---|---|---|---|---|---|---|---|---|---|
| Democratic People's Republic of Korea | 7.7 | 7.7 | 9.6 | 5 | 7.1 | 10.5 | 6.4 | 7.7 | 7 | 7.5 | 6 | 7.5 |
| Democratic Republic of the Congo | 8.1 | 7.7 | 7.2 | 7 | 7.7 | 9.8 | 6.0 | 8.1 | 7 | 7.8 | 8 | 7.8 |
| Denmark | 9.1 | 7.5 | 5.3 | 9 | 7.9 | 5.5 | 6.3 | 10.5 | 10 | 8.7 | 9 | 8.7 |
| Djibouti | 8.3 | 7.8 | 7.8 | 7 | 7.8 | 10.0 | 6.1 | 9.3 | 8 | 8.0 | 8 | 8.1 |
| Dominica | 9.4 | 8.3 | 9.4 | 10 | 8.7 | 12.2 | 9.1 | 11.8 | 11 | 8.8 | 11 | 8.8 |
| Dominican Republic | 9.1 | 8.8 | 10.8 | 9 | 6.7 | 11.1 | 7.7 | 11.2 | 12 | 7.2 | 11 | 6.4 |
| Ecuador | 9.5 | 6.7 | 9.9 | 9 | 6.7 | 11.1 | 8.1 | 10.0 | 9 | 9.7 | 9 | 9.7 |
| Egypt | 10.1 | 7.7 | 8.3 | 9 | 7.6 | 8.9 | 7.4 | 10.5 | 9 | 7.6 | 8 | 7.6 |
| El Salvador | 9.0 | 6.9 | 11 | 8 | 6 | 12.7 | 9.3 | 9.4 | 9 | 6.6 | 8 | 6.7 |
| Equatorial Guinea | 6.6 | 7.6 | 8.7 | 7 | 7.6 | 10.6 | 7.2 | 8.0 | 8 | 7.8 | 8 | 7.9 |
| Eritrea | 7.6 | 5.6 | 7.7 | 5 | 5.3 | 9.9 | 6.5 | 8.7 | 9 | 6.4 | 8 | 6.4 |
| Estonia | 7.6 | 7.1 | 9.3 | 8 | 6.7 | 7.7 | 6.0 | 8.9 | 9 | 7.8 | 9 | 7.3 |
| Ethiopia | 6.0 | 7.6 | 7.1 | 7 | 8 | 10.0 | 6.1 | 8.1 | 8 | 8.6 | 9 | 8.7 |
| Fiji | 8.8 | 7.6 | 8.3 | 9 | 7 | 11.0 | 7.7 | 8.5 | 9 | 7.0 | 9 | 7.0 |
| Finland | 10.0 | 7.6 | 7.6 | 8 | 8.5 | 6.8 | 6.1 | 11.8 | 10 | 9.9 | 10 | 9.8 |
| France | 9.4 | 8.6 | 6.7 | 8 | 9.2 | 6.6 | 6.7 | 10.5 | 10 | 10.2 | 10 | 10.2 |
| Gabon | 8.0 | 8.3 | 7.8 | 9 | 8.2 | 9.8 | 7.1 | 7.6 | 9 | 8.6 | 9 | 8.5 |
| Gambia | 8.1 | 8.3 | 8.6 | 7 | 8.4 | 11.1 | 6.9 | 8.5 | 7 | 8.5 | 8 | 8.5 |
| Georgia | 8.0 | 7.4 | 9.6 | 8 | 7.5 | 7.9 | 6.2 | 8.1 | 8 | 7.6 | 9 | 7.6 |
| Germany | 9.1 | 8.0 | 6.9 | 8 | 8.9 | 6.8 | 5.9 | 10.4 | 9 | 9.8 | 10 | 9.8 |
| Ghana | 8.9 | 9.9 | 8.5 | 8 | 10.5 | 10.0 | 7.2 | 8.7 | 8 | 11.2 | 9 | 10.0 |
| Greece | 9.5 | 8.4 | 5.7 | 9 | 8.7 | 6.5 | 6.7 | 10.1 | 9 | 9.1 | 10 | 9.5 |
| Grenada | 9.3 | 7.3 | 8.8 | 9 | 7 | 9.7 | 7.5 | 11.2 | 9 | 7.3 | 10 | 7.3 |
| Guatemala | 8.6 | 6.8 | 10.1 | 8 | 8.5 | 12.2 | 8.2 | 8.8 | 9 | 8.7 | 8 | 8.7 |
| Guinea | 7.8 | 8.2 | 8.6 | 7 | 8.2 | 10.1 | 7.0 | 8.6 | 8 | 8.4 | 8 | 8.4 |
| Guinea-Bissau | 7.4 | 7.8 | 7.7 | 8 | 8 | 9.8 | 6.1 | 8.1 | 7 | 8.0 | 8 | 8.0 |
| Guyana | 8.5 | 6.0 | 10.1 | 8 | 6.6 | 9.7 | 8.4 | 10.6 | 8 | 7.5 | 8 | 7.5 |
| Haiti | 8.2 | 7.7 | 8.4 | 8 | 7.8 | 7.1 | 5.6 | 4.7 | 11 | 7.9 | 11 | 7.9 |
| Honduras | 9.2 | 8.0 | 10.6 | 10 | 8 | 12.3 | 7.9 | 9.5 | 10 | 8.2 | 10 | 8.2 |
| Hungary | 8.1 | 7.3 | 11 | 10 | 7.3 | 9.3 | 6.8 | 9.3 | 10 | 7.1 | 10 | 7.1 |
| Iceland | 11.5 | 8.5 | 7.3 | 9 | 9.3 | 7.6 | 6.3 | 13.1 | 9 | 11.0 | 10 | 10.3 |
| India | 8.0 | 5.9 | 7.6 | 9 | 6 | 8.4 | 6.8 | 8.3 | 8 | 6.2 | 9 | 6.2 |
| Indonesia | 8.4 | 7.5 | 6.9 | 8 | 7.8 | 8.3 | 7.5 | 8.4 | 8 | 8.1 | 8 | 8.1 |
| Iran, Islamic Republic of | 9.3 | 7.4 | 9.1 | 9 | 8.3 | 10.9 | 10.4 | 10.1 | 9 | 8.5 | 9 | 8.5 |
| Iraq | 9.6 | 8.1 | 9.2 | 8 | 8.2 | 11.0 | 10.3 | 9.8 | 8 | 7.7 | 8 | 7.7 |
| Ireland | 8.8 | 7.6 | 6.3 | 8 | 8.3 | 6.1 | 6.3 | 10.4 | 10 | 9.2 | 10 | 9.2 |
| Israel | 9.6 | 8.5 | 7.3 | 10 | 9.1 | 8.1 | 6.9 | 10.9 | 10 | 9.8 | 10 | 9.7 |
| Italy | 9.2 | 8.2 | 6.4 | 9 | 8.9 | 7.0 | 6.0 | 10.6 | 9 | 9.8 | 9 | 9.9 |
| Jamaica | 9.7 | 8.9 | 10 | 8 | 9 | 9.9 | 6.9 | 12.3 | 11 | 9.4 | 11 | 9.3 |
| Japan | 8.3 | 8.9 | 6.3 | 8 | 9 | 6.5 | 6.1 | 8.7 | 8 | 9.4 | 8 | 9.5 |
| Jordan | 9.8 | 8.1 | 10.3 | 8 | 8.3 | 11.4 | 9.0 | 10.9 | 8 | 8.5 | 8 | 8.6 |
| Kazakhstan | 7.5 | 6.0 | 7.5 | 7 | 6 | 9.8 | 6.1 | 7.4 | 7 | 6.7 | 7 | 6.6 |
| Kenya | 8.6 | 8.3 | 7 | 6 | 8 | 8.7 | 5.7 | 8.5 | 7 | 8.3 | 8 | 8.4 |
| Kiribati | 8.4 | 7.4 | 7.6 | 8 | 7.6 | 10.6 | 9.5 | 8.2 | 8 | 7.9 | 8 | 7.9 |
| Kuwait | 11.0 | 8.5 | 9.6 | 9 | 9.2 | 10.8 | 8.2 | 10.8 | 10 | 9.6 | 10 | 9.6 |
| Kyrgyzstan | 8.2 | 5.7 | 10.4 | 7 | 7.4 | 12.5 | 8.2 | 8.1 | 8 | 6.8 | 8 | 6.8 |
| Lao People's Democratic Republic | 7.3 | 7.1 | 8.6 | 8 | 7.5 | 11.1 | 7.0 | 8.3 | 8 | 7.9 | 9 | 7.9 |
| Latvia | 8.1 | 6.8 | 12.8 | 8 | 6.6 | 10.1 | 6.6 | 8.9 | 8 | 7.0 | 8 | 7.0 |
| Lebanon | 9.1 | 8.0 | 8.9 | 9 | 8.6 | 11.1 | 8.4 | 10.3 | 10 | 9.3 | 9 | 9.4 |
| Lesotho | 7.6 | 7.1 | 5.9 | 6 | 6.7 | 6.9 | 3.3 | 6.4 | 7 | 6.9 | 7 | 6.9 |
| Liberia | 7.4 | 7.5 | 8.4 | 8 | 7.7 | 9.3 | 6.5 | 8.9 | 9 | 8.3 | 9 | 8.3 |
| Libyan Arab Jamahiriya | 10.5 | 8.0 | 9.2 | 9 | 8.3 | 11.4 | 8.1 | 10.7 | 9 | 8.7 | 9 | 8.8 |
| Lithuania | 8.4 | 6.6 | 13.3 | 9 | 6.8 | 10.8 | 7.2 | 8.7 | 8 | 6.6 | 9 | 6.7 |
| Luxembourg | 9.0 | 7.7 | 6.3 | 9 | 8.3 | 6.4 | 6.4 | 11.1 | 10 | 9.1 | 10 | 9.2 |
| Madagascar | 7.9 | 7.1 | 8.5 | 8 | 6.8 | 11.1 | 7.2 | 9.2 | 8 | 7.0 | 9 | 7.0 |

| Country | | | | | | | | | | | | |
|---|---|---|---|---|---|---|---|---|---|---|---|---|
| Malawi | 7.2 | 9.3 | 5.8 | 6 | 9.5 | 6.7 | 4.8 | 7.2 | 9 | 10.8 | 8 | 11.1 |
| Malaysia | 8.9 | 7.1 | 8.6 | 8 | 7.2 | 11.7 | 8.0 | 8.7 | 9 | 7.2 | 9 | 7.2 |
| Maldives | 9.2 | 8.3 | 10.4 | 8 | 6.7 | 14.3 | 7.5 | 10.2 | 10 | 8.7 | 10 | 8.8 |
| Mali | 6.8 | 8.0 | 7.9 | 7 | 8 | 10.5 | 6.4 | 8.1 | 8 | 8.5 | 7 | 8.5 |
| Malta | 9.7 | 8.2 | 6.7 | 9 | 8.5 | 8.2 | 6.2 | 10.4 | 10 | 9.5 | 9 | 9.2 |
| Marshall Islands | 9.4 | 7.5 | 7.9 | 8 | 7.9 | 10.3 | 7.2 | 8.8 | 10 | 8.3 | 10 | 8.3 |
| Mauritania | 9.3 | 8.3 | 9.6 | 9 | 8.4 | 11.4 | 6.9 | 9.8 | 9 | 8.6 | 9 | 8.6 |
| Mauritius | 7.8 | 7.4 | 9.1 | 8 | 7.6 | 11.0 | 8.1 | 8.5 | 8 | 8.1 | 8 | 8.1 |
| Mexico | 7.8 | 8.8 | 7.9 | 8 | 8.9 | 9.0 | 8.3 | 7.8 | 8 | 7.6 | 8 | 8.8 |
| Micronesia, Federated States of | 8.4 | 7.9 | 8 | 9 | 7.9 | 10.6 | 7.9 | 8.2 | 9 | 8.1 | 9 | 8.1 |
| Monaco | | 8.7 | 7.4 | 9 | 8.9 | 7.5 | 7.1 | | 9 | 11.4 | 9 | 11.4 |
| Mongolia | 7.2 | 7.5 | 10.9 | 7 | 7.4 | 11.4 | 6.8 | 7.3 | 8 | 7.3 | 7 | 7.3 |
| Morocco | 10.3 | 8.1 | 10.8 | 9 | 8.2 | 12.6 | 9.4 | 10.6 | 9 | 8.2 | 9 | 8.2 |
| Mozambique | 7.2 | 7.3 | 6.4 | 6 | 7.5 | 9.3 | 4.9 | 7.1 | 7 | 7.9 | 7 | 7.9 |
| Myanmar | | 7.3 | 8.5 | 8 | 7.5 | 8.2 | 6.3 | | 8 | 7.7 | 8 | 7.7 |
| Namibia | 8.3 | 7.4 | 6.3 | 8 | 7.3 | 8.6 | 5.2 | 8.4 | 9 | 7.9 | 10 | 8.1 |
| Nauru | | 7.4 | 8.3 | 11 | 7.8 | 9.9 | 6.9 | | 11 | 8.1 | 11 | 8.1 |
| Nepal | 8.7 | 7.2 | 11 | 8 | 7.6 | 9.9 | 7.4 | 10.1 | 9 | 7.9 | 9 | 8.0 |
| Netherlands | 9.3 | 7.7 | 7.3 | 9 | 9.1 | 7.1 | 6.3 | 10.6 | 9 | 9.8 | 9 | 9.9 |
| New Zealand | 9.7 | 7.9 | 6.4 | 9 | 8.6 | 6.9 | 7.2 | 10.9 | 9 | 9.9 | 9 | 9.9 |
| Nicaragua | 9.4 | 8.4 | 10.6 | 9 | 7.6 | 12.7 | 8.2 | 9.6 | 8 | 7.4 | 9 | 7.4 |
| Niger | 6.6 | 8.0 | 8.8 | 7 | 8.2 | 10.2 | 6.8 | 8.4 | 9 | 8.4 | 8 | 8.4 |
| Nigeria | 8.1 | 7.9 | 7.7 | 7 | 7.9 | 10.6 | 6.8 | 8.8 | 7 | 8.2 | 7 | 8.2 |
| Niue | | 7.8 | 8.7 | 10 | 8 | 11.3 | 8.6 | | 10 | 8.2 | 10 | 8.2 |
| Norway | 10.7 | 8.2 | 6.9 | 10 | 9.2 | 6.8 | 5.9 | 12.2 | 11 | 10.1 | 11 | 9.9 |
| Oman | 9.7 | 7.8 | 10.3 | 9 | 7.7 | 10.4 | 8.3 | 10.2 | 9 | 7.8 | 9 | 7.8 |
| Pakistan | 8.6 | 8.0 | 10 | 9 | 8 | 10.7 | 6.9 | 8.7 | 8 | 8.1 | 9 | 8.1 |
| Palau | | 7.5 | 8.2 | 9 | 7.9 | 11.4 | 7.7 | | 10 | 8.2 | 10 | 8.3 |
| Panama | 9.7 | 8.0 | 8.9 | 10 | 8 | 10.8 | 8.5 | 9.3 | 9 | 8.1 | 9 | 8.2 |
| Papua New Guinea | 7.9 | 5.7 | 8.5 | 7 | 5.9 | 10.5 | 7.0 | 7.9 | 8 | 6.2 | 8 | 6.2 |
| Paraguay | 10.3 | 8.9 | 10.3 | 9 | 7.5 | 12.9 | 9.1 | 9.7 | 9 | 8.6 | 9 | 8.6 |
| Peru | 9.8 | 9.0 | 8.9 | 9 | 8.4 | 11.5 | 7.9 | 10.4 | 10 | 8.9 | 10 | 8.9 |
| Philippines | 9.2 | 7.6 | 7.7 | 9 | 7.8 | 13.1 | 8.0 | 9.2 | 8 | 7.3 | 8 | 7.3 |
| Poland | 8.3 | 7.1 | 10 | 9 | 7.4 | 7.8 | 7.5 | 9.3 | 9 | 7.2 | 10 | 7.3 |
| Portugal | 8.9 | 7.9 | 7.8 | 8 | 8.4 | 8.5 | 6.9 | 9.9 | 9 | 9.0 | 10 | 9.0 |
| Qatar | 11.8 | 9.8 | 11.1 | 11 | 8 | 11.5 | 8.2 | 12.7 | 11 | 7.9 | 11 | 8.0 |
| Republic of Korea | 7.5 | 7.4 | 7.3 | 8 | 7.7 | 6.7 | 6.9 | 8.6 | 8 | 8.9 | 8 | 9.2 |
| Republic of Moldova | 7.8 | 6.8 | 7.7 | 7 | 6.6 | 10.0 | 6.8 | 8.0 | 7 | 6.6 | 7 | 6.7 |
| Romania | 8.2 | 7.5 | 6.8 | 8 | 7.6 | 9.2 | 7.0 | 8.7 | 8 | 7.9 | 8 | 7.8 |
| Russian Federation | 7.7 | 6.6 | 9.1 | 8 | 6.5 | 7.4 | 5.5 | 7.7 | 8 | 6.9 | 8 | 6.9 |
| Rwanda | 6.8 | 7.3 | 6.5 | 7 | 8.1 | 7.3 | 5.6 | 8.8 | 9 | 8.2 | 9 | 8.4 |
| Saint Kitts and Nevis | | 8.2 | 8.4 | 8 | 9 | 10.2 | 8.7 | | 10 | 8.0 | 11 | 8.0 |
| Saint Lucia | 9.5 | 7.8 | 8.5 | 9 | 7 | 10.7 | 8.6 | 11.9 | 11 | 7.2 | 12 | 7.2 |
| Saint Vincent and the Grenadines | 9.3 | 7.8 | 8 | 8 | 7.8 | 10.3 | 7.9 | 11.6 | 11 | 8.2 | 11 | 8.2 |
| Samoa | 8.7 | 6.3 | 8.5 | 8 | 6.8 | 11.0 | 7.6 | 8.6 | 8 | 7.2 | 8 | 7.2 |
| San Marino | | 9.7 | 6.5 | 9 | 10.5 | 7.2 | 6.3 | | 10 | 10.3 | 10 | 10.3 |
| Sao Tome and Principe | 9.0 | 7.9 | 10 | 8 | 8 | 14.8 | 7.5 | 9.7 | 9 | 8.2 | 9 | 8.2 |
| Saudi Arabia | 10.8 | 8.1 | 9.7 | 10 | 8.2 | 10.9 | 8.6 | 11.1 | 10 | 9.1 | 9 | 9.1 |
| Senegal | 7.9 | 8.3 | 8.8 | 8 | 8.3 | 11.3 | 7.3 | 8.7 | 9 | 8.5 | 8 | 8.5 |
| Seychelles | 7.4 | 6.8 | 9.5 | 6 | 8.7 | 11.3 | 9.6 | 7.1 | 6 | 7.4 | 7 | 7.5 |
| Sierra Leone | 7.2 | 7.1 | 7.3 | 5 | 7 | 8.6 | 5.1 | 8.9 | 6 | 7.4 | 7 | 7.5 |
| Singapore | 8.0 | 8.1 | 8.6 | 6 | 8.4 | 8.6 | 8.6 | 9.2 | 6 | 8.7 | 6 | 9.7 |
| Slovakia | 8.5 | 7.3 | 9.7 | 9 | 7.4 | 7.7 | 6.7 | 9.2 | 9 | 7.3 | 9 | 7.4 |
| Slovenia | 8.6 | 7.9 | 7.4 | 9 | 8.2 | 7.0 | 6.1 | 10.2 | 11 | 8.2 | 11 | 8.3 |
| Solomon Islands | 7.9 | 7.3 | 8.6 | 9 | 7.6 | 12.0 | 8.3 | 7.5 | 9 | 7.9 | 8 | 7.9 |

| Country | HALE 1990 | Model 1990 | HALEb 2000 | HALE 2000 | Model 2000 | HALE 2001 | HALE 2002 | HALE 2010 | HALE 2012 | Model 2012 | HALE 2013 | Model 2013 |
|---|---|---|---|---|---|---|---|---|---|---|---|---|
| Somalia | 7.1 | 6.9 | 8.3 | 7 | 7 | 8.5 | 6.9 | 7.8 | 7 | 7.1 | 8 | 7.2 |
| South Africa | 8.1 | 6.4 | 6.6 | 7 | 6.5 | 7.7 | 5.5 | 8.3 | 7 | 6.7 | 8 | 6.7 |
| Spain | 8.5 | 8.1 | 6.6 | 8 | 8.5 | 6.6 | 6.2 | 9.6 | 8 | 9.1 | 9 | 9.0 |
| Sri Lanka | 9.4 | 6.3 | 9 | 8 | 6.1 | 11.5 | 8.0 | 9.3 | 8 | 8.0 | 9 | 7.9 |
| Sudan | 10.1 | 7.2 | 9.8 | 9 | 7.2 | 11.2 | 7.8 | 11.0 | 9 | 7.4 | 9 | 7.4 |
| Suriname | 9.1 | 8.9 | 8.5 | 10 | 9.9 | 10.0 | 7.6 | 11.6 | 12 | 10.9 | 11 | 11.1 |
| Swaziland | 8.4 | 7.6 | 6 | 7 | 7.3 | 6.4 | 3.7 | 7.0 | 8 | 7.9 | 8 | 7.8 |
| Sweden | 9.9 | 8.6 | 7.2 | 10 | 9.5 | 7.2 | 6.2 | 11.2 | 10 | 10.4 | 10 | 10.4 |
| Switzerland | 9.1 | 8.8 | 6.2 | 9 | 9.8 | 6.2 | 6.6 | 10.6 | 11 | 10.3 | 10 | 10.4 |
| Syrian Arab Republic | 10.1 | 8.3 | 9.7 | 10 | 8.5 | 10.7 | 8.5 | 10.5 | 7 | 8.5 | 8 | 8.7 |
| Tajikistan | 8.2 | 7.7 | 10.8 | 7 | 7.7 | 12.8 | 7.9 | 8.7 | 8 | 7.8 | 8 | 7.8 |
| Thailand | 8.5 | 7.2 | 8.4 | 8 | 8 | 9.3 | 8.4 | 8.2 | 8 | 8.6 | 8 | 8.7 |
| The former Yugoslav Republic of Macedonia | 8.9 | 7.7 | 6.3 | 9 | 7.4 | 8.5 | 7.2 | 9.6 | 9 | 8.1 | 10 | 8.2 |
| Togo | 8.4 | 9.3 | 7.9 | 8 | 10.3 | 9.7 | 6.5 | 8.3 | 8 | 10.7 | 8 | 10.7 |
| Tonga | 9.7 | 6.9 | 8.1 | 8 | 7.2 | 11.0 | 8.2 | 8.4 | 10 | 7.9 | 10 | 8.0 |
| Trinidad and Tobago | 8.6 | 6.5 | 8.2 | 8 | 6.5 | 8.4 | 7.3 | 10.5 | 9 | 6.6 | 9 | 6.6 |
| Tunisia | 9.0 | 8.5 | 8.2 | 8 | 8.5 | 10.1 | 8.2 | 9.5 | 9 | 8.5 | 9 | 8.6 |
| Turkey | 8.4 | 8.4 | 10 | 9 | 8.9 | 8.5 | 6.7 | 9.4 | 9 | 9.3 | 9 | 9.3 |
| Turkmenistan | 7.5 | 6.9 | 8.8 | 6 | 6.9 | 12.1 | 7.1 | 8.3 | 7 | 7.2 | 7 | 7.2 |
| Tuvalu |  | 6.1 | 7.2 | 9 | 6.3 | 9.9 | 7.0 |  | 9 | 6.7 | 9 | 6.7 |
| Uganda | 7.6 | 7.4 | 7.2 | 6 | 7.3 | 9.0 | 6.2 | 8.2 | 8 | 7.9 | 8 | 8.0 |
| Ukraine | 8.0 | 6.9 | 10.3 | 7 | 6.8 | 9.3 | 6.8 | 7.9 | 7 | 7.1 | 7 | 7.1 |
| United Arab Emirates | 10.1 | 7.6 | 10 | 9 | 7.9 | 9.0 | 7.8 | 10.6 | 10 | 8.3 | 9 | 8.4 |
| United Kingdom | 9.5 | 7.6 | 6.5 | 10 | 8.6 | 6.6 | 6.7 | 10.7 | 10 | 9.2 | 10 | 9.2 |
| United Republic of Tanzania | 8.3 | 7.8 | 7.2 | 7 | 7.9 | 9.5 | 5.5 | 9.1 | 8 | 8.3 | 9 | 8.3 |
| United States of America | 8.7 | 7.6 | 8.2 | 8 | 8.5 | 8.0 | 7.4 | 9.7 | 9 | 9.4 | 8 | 9.4 |
| Uruguay | 8.3 | 8.3 | 8.4 | 8 | 6.7 | 9.7 | 8.0 | 8.6 | 8 | 7.2 | 9 | 7.2 |
| Uzbekistan | 8.4 | 7.5 | 9.4 | 7 | 7.3 | 11.7 | 7.6 | 8.5 | 8 | 7.1 | 8 | 7.1 |
| Vanuatu | 8.1 | 7.6 | 8.2 | 9 | 7.9 | 11.0 | 8.0 | 7.9 | 9 | 8.2 | 10 | 8.2 |
| Venezuela, Bolivarian Republic of | 8.9 | 7.6 | 10.1 | 9 | 7.9 | 13.7 | 9.3 | 8.6 | 9 | 9.3 | 9 | 9.3 |
| Viet Nam | 8.8 | 6.8 | 8.5 | 9 | 6.7 | 11.0 | 7.4 | 9.0 | 9 | 6.8 | 9 | 6.8 |
| Yemen | 9.5 | 7.4 | 10.4 | 8 | 7.5 | 12.9 | 10.8 | 10.2 | 8 | 7.6 | 9 | 7.6 |
| Zambia | 7.3 | 6.7 | 5.5 | 5 | 6.8 | 6.2 | 4.3 | 7.5 | 7 | 7.8 | 8 | 7.9 |
| Zimbabwe | 9.6 | 8.7 | 5.8 | 5 | 6.7 | 5.6 | 3.9 | 7.8 | 9 | 8.7 | 8 | 8.9 |
| Method | HALE | Model | HALEb | HALE | Model | HALE | HALE | HALE | HALE | Model | HALE | Model |
| Year | 1990 | 1990 | 2000 | 2000 | 2000 | 2001 | 2002 | 2010 | 2012 | 2012 | 2013 | 2013 |
| Mean | 8.67 | **7.69** | 8.41 | 8.18 | **7.83** | 9.73 | 7.19 | 9.33 | 8.79 | **8.25** | 8.88 | 8.28 |

TABLE VI

Estimates of the Mortality Model for the Loss of Healthy Life Years (LHLY) for females for the WHO member countries and the related results from the HALE method of the World Health Organization

| Countries / Year | HALE 1990 | Model 1990 | HALE 2000 | HALE 2000b | Model 2000 | HALE 2001 | HALE 2002 | HALE 2010 | HALE 2012 | Model 2012 | HALE 2013 | Model 2013 |
|---|---|---|---|---|---|---|---|---|---|---|---|---|
| Afghanistan | 10.1 | 7.1 | 12.5 | 11.0 | 7.5 | 8.1 | 7.7 | 11.1 | 12 | 7.9 | 12 | 8 |
| Albania | 11.0 | 11.3 | 10.6 | 9.0 | 10.4 | 11.7 | 10.8 | 11.1 | 9 | 10.1 | 10 | 10 |
| Algeria | 11.2 | 8.9 | 12.9 | 10.0 | 8.9 | 11.2 | 9.6 | 11.9 | 10 | 8.9 | 11 | 9 |
| Andorra | 12.5 | 11.1 | 10.1 | 11.0 | 11.3 | 10.0 | 9.1 | 13.0 | 12 | 12.5 | 12 | 13 |
| Angola | 8.3 | 7.4 | 10.8 | 7.0 | 7.6 | 6.5 | 6.9 | 9.9 | 7 | 7.9 | 7 | 8 |
| Antigua and Barbuda | 11.2 | 9.5 | 14.5 | 9.0 | 9.2 | 10.9 | 10.3 | 13.5 | 11 | 9.7 | 11 | 10 |
| Argentina | 10.2 | 9.0 | 11.9 | 10.0 | 8.8 | 12.0 | 10.0 | 10.6 | 10 | 9.5 | 11 | 10 |
| Armenia | 10.7 | 7.9 | 10.1 | 10.0 | 8.3 | 11.9 | 10.4 | 11.3 | 9 | 8.9 | 9 | 9 |
| Australia | 11.2 | 9.7 | 8.8 | 11.0 | 10.4 | 9.5 | 8.7 | 12.0 | 11 | 11.6 | 11 | 12 |
| Austria | 10.7 | 10.4 | 8.9 | 10.0 | 11.1 | 8.8 | 8.6 | 12.1 | 11 | 11.8 | 11 | 12 |
| Azerbaijan | 10.5 | 7.8 | 11.4 | 9.0 | 7.8 | 11.2 | 10.0 | 11.1 | 10 | 8.6 | 10 | 9 |
| Bahamas | 12.2 | 9.7 | 15.7 | 11.0 | 11.3 | 12.5 | 9.7 | 13.9 | 12 | 10.8 | 11 | 11 |
| Bahrain | 9.9 | 7.9 | 12.4 | 12.0 | 7.6 | 12.2 | 10.1 | 11.2 | 12 | 9.6 | 12 | 10 |
| Bangladesh | 11.5 | 7.9 | 12.9 | 10.0 | 7.8 | 9.2 | 9.3 | 12.3 | 10 | 7.7 | 10 | 8 |
| Barbados | 10.4 | 8.4 | 13.4 | 12.0 | 8.7 | 10.6 | 9.8 | 10.4 | 13 | 10.2 | 13 | 10 |
| Belarus | 11.2 | 8.4 | 9.2 | 10.0 | 8.1 | 11.4 | 9.4 | 11.7 | 10 | 9.1 | 10 | 9 |
| Belgium | 11.7 | 10.2 | 9.9 | 10.0 | 10.6 | 9.4 | 8.2 | 12.1 | 11 | 11.4 | 11 | 11 |
| Belize | 10.3 | 9.8 | 14.3 | 11.0 | 9.5 | 11.3 | 10.2 | 10.8 | 12 | 9.8 | 12 | 10 |
| Benin | 9.2 | 8.5 | 11.9 | 9.0 | 8.3 | 9.2 | 7.9 | 10.2 | 9 | 8.6 | 9 | 9 |
| Bhutan | 9.4 | 6.0 | 14.3 | 8.0 | 6.1 | 9.9 | 9.5 | 10.2 | 10 | 6.4 | 9 | 6 |
| Bolivia | 11.1 | 8.5 | 12.1 | 9.0 | 8.8 | 10.7 | 9.4 | 10.7 | 9 | 8.6 | 9 | 9 |
| Bosnia and Herzegovina | 10.8 | 8.8 | 9.4 | 10.0 | 7.1 | 11.5 | 10.0 | 12.7 | 11 | 9.3 | 10 | 9 |
| Botswana | 10.7 | 7.9 | 7.9 | 6.0 | 7.4 | 5.9 | 5.2 | 11.1 | 10 | 7.5 | 10 | 8 |
| Brazil | 10.0 | 8.2 | 12.7 | 10.0 | 7.7 | 11.0 | 9.8 | 10.5 | 10 | 8.1 | 11 | 7 |
| Brunei Darussalam | 9.8 | 8.5 | 12.7 | 10.0 | 8.0 | 12.2 | 11.9 | 10.2 | 9 | 9.3 | 10 | 9 |
| Bulgaria | 8.9 | 9.3 | 9.2 | 9.0 | 9.5 | 9.6 | 8.5 | 8.8 | 10 | 10.0 | 10 | 10 |
| Burkina Faso | 8.5 | 8.2 | 9.5 | 7.0 | 8.3 | 7.2 | 6.3 | 9.3 | 8 | 8.4 | 8 | 8 |
| Burundi | 7.7 | 7.2 | 8.5 | 7.0 | 7.3 | 6.6 | 6.2 | 8.4 | 8 | 7.6 | 9 | 8 |
| Cambodia | 9.6 | 6.6 | 9.8 | 10.0 | 7.0 | 9.1 | 7.6 | 10.1 | 12 | 8.0 | 12 | 8 |
| Cameroon | 9.9 | 8.0 | 10.5 | 8.0 | 7.8 | 8.4 | 7.3 | 9.7 | 12 | 8.1 | 11 | 8 |
| Canada | 11.1 | 9.5 | 9.8 | 10.0 | 10.3 | 10.4 | 8.3 | 11.8 | 8 | 10.7 | 9 | 11 |
| Cape Verde | 12.0 | 8.4 | 12.3 | 12.0 | 9.0 | 11.3 | 10.0 | 12.7 | 11 | 9.7 | 11 | 10 |
| Central African Republic | 8.6 | 7.6 | 8.9 | 7.0 | 7.2 | 7.7 | 6.1 | 7.6 | 8 | 7.9 | 8 | 8 |
| Chad | 9.2 | 7.4 | 11.2 | 7.0 | 7.4 | 8.7 | 7.6 | 9.2 | 8 | 7.6 | 8 | 8 |
| Chile | 10.2 | 8.7 | 12.1 | 10.0 | 8.7 | 11.7 | 10.3 | 10.5 | 11 | 9.2 | 11 | 9 |
| China | 8.0 | 8.7 | 9.7 | 7.0 | 9.7 | 8.4 | 7.6 | 8.6 | 8 | 10.1 | 8 | 10 |
| Colombia | 11.0 | 8.7 | 11.8 | 12.0 | 7.5 | 12.7 | 10.0 | 11.2 | 14 | 8.4 | 12 | 11 |
| Comoros | 8.6 | 7.6 | 12.3 | 8.0 | 7.7 | 11.0 | 9.6 | 9.3 | 9 | 7.9 | 8 | 8 |
| Congo | 9.8 | 8.1 | 10.1 | 8.0 | 7.9 | 8.7 | 7.2 | 10.0 | 9 | 8.4 | 9 | 8 |
| Cook Islands |  | 7.4 | 11.0 | 11.0 | 8.1 | 11.4 | 11.5 |  | 12 | 10.1 | 12 | 10 |
| Costa Rica | 11.4 | 9.2 | 12.4 | 10.0 | 8.8 | 11.6 | 10.3 | 11.4 | 10 | 9.2 | 10 | 9 |
| Côte d'Ivoire | 10.1 | 8.0 | 9.5 | 7.0 | 7.4 | 7.7 | 6.7 | 9.6 | 8 | 7.8 | 8 | 8 |
| Croatia | 11.3 | 9.3 | 10.6 | 10.0 | 9.6 | 10.2 | 9.3 | 11.6 | 11 | 9.9 | 11 | 10 |
| Cuba | 11.0 | 10.1 | 10.9 | 11.0 | 8.2 | 10.8 | 9.8 | 12.9 | 13 | 8.7 | 12 | 9 |
| Cyprus | 11.9 | 11.6 | 12.7 | 8.0 | 12.6 | 12.0 | 10.6 | 12.3 | 8 | 12.7 | 8 | 13 |
| Czech Republic | 10.3 | 9.2 | 9.9 | 9.0 | 9.8 | 9.3 | 8.1 | 11.1 | 10 | 10.6 | 10 | 11 |

| Country | | | | | | | | | | | |
|---|---|---|---|---|---|---|---|---|---|---|---|
| Democratic People's Republic of Korea | 8.7 | 9.8 | 11.2 | 8.0 | 9.2 | 10.3 | 7.4 | 8.9 | 8 | 9.6 | 8 | 10 |
| Democratic Republic of the Congo | 9.2 | 7.7 | 9.6 | 8.0 | 7.7 | 8.2 | 7.0 | 9.6 | 8 | 7.9 | 8 | 8 |
| Denmark | 10.6 | 9.1 | 8.4 | 9.0 | 8.9 | 8.7 | 8.4 | 11.5 | 11 | 9.6 | 11 | 10 |
| Djibouti | 9.6 | 8.2 | 10.1 | 9.0 | 8.2 | 8.1 | 7.4 | 10.3 | 10 | 8.5 | 9 | 9 |
| Dominica | 10.7 | 9.3 | 12.2 | 10.0 | 9.7 | 11.2 | 10.2 | 12.9 | 12 | 9.1 | 12 | 9 |
| Dominican Republic | 10.7 | 9.8 | 14.0 | 10.0 | 7.6 | 10.7 | 9.6 | 11.8 | 12 | 8.5 | 10 | 7 |
| Ecuador | 11.1 | 7.4 | 12.0 | 11.0 | 7.4 | 10.8 | 9.4 | 11.3 | 10 | 10.5 | 11 | 11 |
| Egypt | 12.0 | 8.2 | 12.0 | 11.0 | 8.2 | 10.8 | 8.8 | 12.6 | 11 | 8.3 | 11 | 8 |
| El Salvador | 11.2 | 7.9 | 13.9 | 10.0 | 7.1 | 11.5 | 10.4 | 11.2 | 11 | 7.1 | 11 | 7 |
| Equatorial Guinea | 8.7 | 7.6 | 11.4 | 8.0 | 7.6 | 9.2 | 8.5 | 10.7 | 10 | 7.9 | 9 | 8 |
| Eritrea | 9.1 | 6.2 | 10.4 | 8.0 | 6.7 | 9.1 | 8.6 | 10.0 | 11 | 7.5 | 10 | 8 |
| Estonia | 10.0 | 8.9 | 11.0 | 9.0 | 9.2 | 10.4 | 8.1 | 11.3 | 10 | 10.3 | 11 | 10 |
| Ethiopia | 6.9 | 7.6 | 9.6 | 8.0 | 8.0 | 8.5 | 7.7 | 8.8 | 9 | 9.1 | 9 | 9 |
| Fiji | 9.9 | 8.4 | 10.7 | 10.0 | 8.0 | 11.0 | 9.7 | 9.8 | 11 | 8.1 | 10 | 8 |
| Finland | 12.5 | 9.8 | 9.5 | 10.0 | 10.9 | 8.8 | 8.0 | 13.7 | 11 | 11.8 | 11 | 12 |
| France | 11.7 | 10.9 | 10.2 | 11.0 | 11.4 | 9.5 | 8.8 | 12.4 | 11 | 12.3 | 11 | 12 |
| Gabon | 10.6 | 8.6 | 10.4 | 9.0 | 8.5 | 9.0 | 8.8 | 10.5 | 10 | 8.8 | 10 | 9 |
| Gambia | 9.7 | 8.4 | 12.1 | 9.0 | 8.6 | 10.1 | 8.4 | 9.8 | 10 | 8.9 | 9 | 9 |
| Georgia | 10.5 | 8.8 | 11.6 | 10.0 | 8.9 | 10.2 | 8.4 | 11.0 | 10 | 9.2 | 10 | 9 |
| Germany | 10.9 | 10.0 | 9.2 | 10.0 | 10.5 | 8.9 | 7.6 | 11.9 | 10 | 11.6 | 10 | 12 |
| Ghana | 10.5 | 11.2 | 11.0 | 9.0 | 10.5 | 9.2 | 8.5 | 10.6 | 10 | 11.2 | 9 | 11 |
| Greece | 11.4 | 10.1 | 8.5 | 10.0 | 11.0 | 8.9 | 8.1 | 11.7 | 10 | 12.2 | 11 | 12 |
| Grenada | 10.9 | 8.8 | 11.5 | 11.0 | 9.3 | 9.7 | 8.9 | 11.8 | 12 | 9.9 | 11 | 10 |
| Guatemala | 9.8 | 7.2 | 12.6 | 10.0 | 8.9 | 11.9 | 9.1 | 10.2 | 10 | 9.1 | 10 | 9 |
| Guinea | 8.9 | 8.3 | 11.9 | 9.0 | 8.1 | 9.1 | 8.2 | 9.9 | 9 | 8.5 | 9 | 9 |
| Guinea-Bissau | 8.7 | 8.2 | 10.5 | 8.0 | 8.1 | 8.2 | 7.2 | 9.1 | 9 | 8.2 | 8 | 8 |
| Guyana | 10.5 | 6.6 | 14.2 | 9.0 | 7.8 | 10.0 | 9.7 | 11.5 | 10 | 7.8 | 10 | 8 |
| Haiti | 8.7 | 8.0 | 11.2 | 8.0 | 8.3 | 7.4 | 6.9 | 6.5 | 11 | 8.2 | 11 | 8 |
| Honduras | 11.0 | 8.8 | 13.2 | 11.0 | 8.9 | 10.7 | 9.9 | 11.0 | 12 | 9.1 | 12 | 9 |
| Hungary | 10.4 | 8.8 | 10.7 | 10.0 | 9.2 | 10.5 | 8.6 | 11.1 | 11 | 9.5 | 11 | 10 |
| Iceland | 12.8 | 9.8 | 9.3 | 10.0 | 10.7 | 9.4 | 8.2 | 14.5 | 11 | 12.1 | 11 | 12 |
| India | 9.2 | 6.4 | 11.0 | 10.0 | 6.7 | 10.4 | 8.4 | 9.8 | 10 | 7.1 | 9 | 7 |
| Indonesia | 9.2 | 8.1 | 9.1 | 9.0 | 8.4 | 10.1 | 9.1 | 9.3 | 9 | 8.8 | 9 | 9 |
| Iran, Islamic Republic of | 11.5 | 7.6 | 11.4 | 10.0 | 8.9 | 13.2 | 12.5 | 12.5 | 11 | 9.8 | 11 | 10 |
| Iraq | 10.5 | 8.9 | 12.1 | 11.0 | 9.1 | 9.6 | 11.6 | 10.5 | 11 | 9.1 | 10 | 9 |
| Ireland | 10.4 | 9.1 | 8.8 | 9.0 | 9.5 | 8.9 | 8.2 | 11.7 | 10 | 10.1 | 10 | 10 |
| Israel | 10.8 | 9.4 | 10.0 | 10.0 | 10.2 | 10.0 | 9.0 | 12.0 | 11 | 11.2 | 10 | 11 |
| Italy | 11.2 | 10.2 | 9.6 | 11.0 | 10.8 | 9.3 | 7.8 | 12.0 | 11 | 11.9 | 11 | 12 |
| Jamaica | 11.0 | 9.3 | 11.5 | 11.0 | 9.7 | 10.0 | 8.6 | 12.7 | 11 | 10.1 | 11 | 10 |
| Japan | 9.7 | 10.9 | 8.4 | 9.0 | 11.1 | 8.9 | 7.5 | 10.4 | 10 | 11.7 | 9 | 12 |
| Jordan | 11.3 | 8.7 | 13.6 | 10.0 | 8.9 | 13.6 | 10.9 | 11.9 | 10 | 9.2 | 11 | 9 |
| Kazakhstan | 10.0 | 7.3 | 10.3 | 8.0 | 7.5 | 11.3 | 9.6 | 9.8 | 8 | 9.3 | 9 | 9 |
| Kenya | 9.7 | 8.5 | 9.4 | 7.0 | 8.4 | 7.5 | 7.1 | 10.1 | 8 | 8.7 | 9 | 9 |
| Kiribati | 9.3 | 8.0 | 10.1 | 10.0 | 8.3 | 10.5 | 11.0 | 10.3 | 9 | 8.7 | 9 | 9 |
| Kuwait | 12.6 | 9.7 | 12.0 | 11.0 | 9.8 | 10.2 | 10.6 | 12.6 | 12 | 10.0 | 12 | 10 |
| Kyrgyzstan | 10.3 | 7.1 | 13.2 | 9.0 | 8.9 | 12.8 | 10.6 | 10.5 | 9 | 8.3 | 9 | 8 |
| Lao People's Democratic Republic | 8.4 | 7.4 | 10.4 | 9.0 | 7.9 | 9.6 | 9.2 | 9.3 | 9 | 8.5 | 10 | 9 |
| Latvia | 10.5 | 8.7 | 11.6 | 10.0 | 9.2 | 11.1 | 8.3 | 11.3 | 11 | 9.6 | 10 | 10 |
| Lebanon | 10.6 | 8.8 | 12.2 | 10.0 | 9.3 | 9.8 | 10.4 | 11.4 | 11 | 10.1 | 11 | 10 |
| Lesotho | 10.1 | 7.4 | 7.7 | 6.0 | 7.2 | 6.3 | 5.0 | 8.1 | 8 | 7.2 | 8 | 7 |
| Liberia | 8.9 | 7.9 | 11.7 | 8.0 | 8.2 | 8.3 | 6.7 | 10.0 | 10 | 8.7 | 10 | 9 |
| Libyan Arab Jamahiriya | 12.5 | 8.8 | 12.4 | 12.0 | 9.1 | 10.8 | 10.5 | 12.9 | 12 | 9.5 | 12 | 9 |
| Lithuania | 10.6 | 9.0 | 14.0 | 10.0 | 9.5 | 12.6 | 9.9 | 10.9 | 11 | 9.9 | 9 | 10 |
| Luxembourg | 11.2 | 9.4 | 8.7 | 10.0 | 9.9 | 9.0 | 8.0 | 12.3 | 11 | 10.4 | 11 | 10 |
| Madagascar | 9.0 | 7.3 | 12.0 | 9.0 | 7.1 | 9.7 | 8.4 | 10.4 | 9 | 7.5 | 10 | 8 |

| Country | | | | | | | | | | | | |
|---|---|---|---|---|---|---|---|---|---|---|---|---|
| Malawi | 8.2 | 8.9 | 7.4 | 6.0 | 8.6 | 6.3 | 5.8 | 8.5 | 9 | 10.0 | 9 | 10 |
| Malaysia | 10.1 | 8.2 | 10.7 | 10.0 | 8.9 | 11.2 | 10.0 | 10.1 | 10 | 9.1 | 10 | 9 |
| Maldives | 9.7 | 8.1 | 13.8 | 9.0 | 8.3 | 10.1 | 9.0 | 11.5 | 10 | 9.8 | 11 | 10 |
| Mali | 8.2 | 8.0 | 10.5 | 8.0 | 8.0 | 8.5 | 7.4 | 9.3 | 9 | 8.5 | 9 | 9 |
| Malta | 11.6 | 10.9 | 8.6 | 10.0 | 10.6 | 9.5 | 8.3 | 12.4 | 11 | 11.3 | 10 | 12 |
| Marshall Islands | 10.7 | 8.0 | 10.4 | 9.0 | 8.5 | 9.6 | 8.9 | 10.2 | 11 | 9.1 | 12 | 9 |
| Mauritania | 10.5 | 8.7 | 12.7 | 11.0 | 8.8 | 9.5 | 8.2 | 10.7 | 11 | 9.0 | 10 | 9 |
| Mauritius | 9.5 | 8.2 | 12.2 | 10.0 | 8.8 | 17.2 | 10.9 | 10.1 | 10 | 9.5 | 10 | 9 |
| Mexico | 9.4 | 9.8 | 10.9 | 9.0 | 9.9 | 11.8 | 9.3 | 9.3 | 10 | 8.3 | 9 | 10 |
| Micronesia, Federated States of | 9.3 | 8.3 | 10.3 | 10.0 | 8.4 | 10.3 | 9.6 | 9.7 | 10 | 8.6 | 9 | 9 |
| Monaco | | 11.2 | 10.5 | 11.0 | 10.9 | 10.5 | 9.3 | | 11 | 13.3 | 11 | 13 |
| Mongolia | 8.3 | 8.6 | 12.4 | 8.0 | 8.1 | 10.3 | 8.0 | 9.0 | 9 | 8.4 | 8 | 8 |
| Morocco | 12.0 | 8.7 | 16.0 | 11.0 | 8.8 | 15.5 | 11.9 | 12.5 | 12 | 8.9 | 11 | 9 |
| Mozambique | 9.1 | 7.5 | 8.4 | 8.0 | 7.6 | 8.3 | 6.4 | 8.8 | 8 | 7.9 | 8 | 8 |
| Myanmar | | 7.8 | 10.7 | 9.0 | 8.0 | 8.5 | 8.4 | | 10 | 8.3 | 9 | 8 |
| Namibia | 10.2 | 8.0 | 7.9 | 8.0 | 6.7 | 8.0 | 6.7 | 9.8 | 10 | 8.1 | 10 | 8 |
| Nauru | | 8.1 | 11.1 | 13.0 | 9.2 | 9.6 | 9.0 | | 14 | 9.6 | 14 | 10 |
| Nepal | 9.1 | 7.4 | 13.8 | 9.0 | 7.9 | 8.8 | 9.1 | 10.7 | 9 | 8.5 | 10 | 9 |
| Netherlands | 11.6 | 9.1 | 9.7 | 11.0 | 10.2 | 9.6 | 8.5 | 12.4 | 11 | 11.3 | 11 | 11 |
| New Zealand | 11.0 | 10.3 | 8.9 | 10.0 | 10.7 | 9.4 | 9.0 | 12.0 | 11 | 11.4 | 11 | 11 |
| Nicaragua | 11.0 | 8.2 | 13.0 | 11.0 | 7.2 | 10.7 | 9.3 | 11.2 | 10 | 7.5 | 11 | 8 |
| Niger | 7.8 | 8.2 | 11.5 | 8.0 | 8.3 | 8.5 | 7.5 | 9.3 | 9 | 8.5 | 8 | 9 |
| Nigeria | 9.2 | 7.9 | 10.3 | 7.0 | 7.8 | 8.9 | 7.8 | 9.6 | 8 | 8.2 | 8 | 8 |
| Niue | | 8.9 | 11.4 | 11.0 | 9.2 | 11.6 | 11.3 | | 12 | 9.4 | 12 | 9 |
| Norway | 12.6 | 9.7 | 9.1 | 12.0 | 10.7 | 9.3 | 8.1 | 13.4 | 12 | 11.2 | 12 | 11 |
| Oman | 11.8 | 7.6 | 13.2 | 12.0 | 7.3 | 12.9 | 11.1 | 12.5 | 11 | 7.1 | 12 | 7 |
| Pakistan | 9.3 | 8.4 | 14.7 | 10.0 | 8.5 | 10.0 | 9.3 | 9.8 | 9 | 8.5 | 10 | 8 |
| Palau | | 8.0 | 10.4 | 10.0 | 8.5 | 10.7 | 10.4 | | 11 | 9.1 | 11 | 9 |
| Panama | 11.6 | 11.3 | 11.0 | 11.0 | 8.9 | 11.1 | 10.2 | 11.2 | 11 | 9.0 | 10 | 9 |
| Papua New Guinea | 8.3 | 6.5 | 10.4 | 9.0 | 6.7 | 9.5 | 9.1 | 8.8 | 10 | 6.9 | 10 | 7 |
| Paraguay | 11.5 | 9.6 | 12.3 | 12.0 | 8.8 | 11.0 | 10.5 | 11.2 | 11 | 9.7 | 11 | 10 |
| Peru | 10.7 | 9.1 | 11.8 | 10.0 | 8.6 | 10.8 | 9.6 | 11.0 | 11 | 9.4 | 11 | 9 |
| Philippines | 11.0 | 8.4 | 10.2 | 11.0 | 9.1 | 11.7 | 10.2 | 10.6 | 9 | 8.8 | 9 | 9 |
| Poland | 10.5 | 9.7 | 13.4 | 10.0 | 9.6 | 11.5 | 10.2 | 11.2 | 11 | 9.8 | 10 | 10 |
| Portugal | 10.9 | 9.9 | 10.7 | 10.0 | 10.4 | 10.7 | 8.8 | 11.6 | 11 | 10.9 | 11 | 11 |
| Qatar | 13.3 | 9.2 | 13.2 | 12.0 | 7.6 | 11.2 | 10.0 | 14.7 | 14 | 7.9 | 13 | 8 |
| Republic of Korea | 9.1 | 8.9 | 9.5 | 9.0 | 10.1 | 8.4 | 8.6 | 10.1 | 10 | 11.6 | 10 | 12 |
| Republic of Moldova | 9.8 | 7.7 | 8.9 | 9.0 | 7.7 | 10.9 | 9.2 | 10.0 | 9 | 8.0 | 9 | 8 |
| Romania | 9.8 | 8.7 | 9.5 | 9.0 | 8.4 | 11.2 | 9.7 | 10.3 | 9 | 9.3 | 9 | 9 |
| Russian Federation | 10.3 | 8.4 | 11.4 | 9.0 | 8.3 | 10.4 | 7.7 | 10.2 | 9 | 8.8 | 9 | 9 |
| Rwanda | 8.1 | 7.4 | 8.7 | 7.0 | 8.3 | 6.8 | 6.6 | 10.7 | 10 | 8.7 | 10 | 9 |
| Saint Kitts and Nevis | | 9.5 | 10.5 | 10.0 | 10.4 | 10.0 | 9.1 | | 12 | 9.5 | 12 | 10 |
| Saint Lucia | 11.1 | 7.8 | 10.9 | 12.0 | 7.2 | 10.6 | 10.2 | 12.4 | 13 | 8.0 | 13 | 8 |
| Saint Vincent and the Grenadines | 11.1 | 8.8 | 11.3 | 10.0 | 8.8 | 10.2 | 9.8 | 12.0 | 11 | 10.0 | 11 | 10 |
| Samoa | 9.9 | 7.3 | 11.3 | 10.0 | 7.8 | 10.4 | 9.4 | 10.2 | 11 | 8.2 | 10 | 8 |
| San Marino | | 12.0 | 9.5 | 11.0 | 12.2 | 9.8 | 8.1 | | 11 | 13.6 | 11 | 14 |
| Sao Tome and Principe | 10.3 | 8.1 | 12.2 | 10.0 | 8.2 | 10.3 | 9.0 | 11.5 | 11 | 8.4 | 10 | 8 |
| Saudi Arabia | 12.9 | 9.6 | 12.8 | 12.0 | 10.0 | 11.0 | 11.0 | 13.3 | 12 | 9.2 | 12 | 9 |
| Senegal | 9.9 | 8.7 | 11.6 | 10.0 | 8.7 | 9.5 | 8.4 | 10.6 | 10 | 9.0 | 10 | 9 |
| Seychelles | 9.6 | 9.0 | 13.8 | 8.0 | 10.1 | 13.6 | 12.3 | 9.1 | 7 | 9.6 | 7 | 10 |
| Sierra Leone | 8.9 | 7.4 | 9.6 | 6.0 | 7.4 | 6.9 | 5.8 | 10.2 | 7 | 7.6 | 6 | 8 |
| Singapore | 9.7 | 9.1 | 11.3 | 8.0 | 9.8 | 11.6 | 10.4 | 10.7 | 8 | 9.4 | 7 | 11 |
| Slovakia | 11.1 | 9.3 | 12.3 | 9.0 | 9.4 | 10.7 | 8.9 | 10.8 | 10 | 10.0 | 10 | 10 |
| Slovenia | 11.0 | 10.0 | 10.2 | 11.0 | 10.3 | 9.2 | 8.2 | 11.8 | 11 | 11.1 | 12 | 11 |
| Solomon Islands | 8.6 | 7.5 | 11.3 | 10.0 | 8.0 | 11.5 | 10.3 | 8.7 | 10 | 8.5 | 9 | 8 |

| Country | | | | | | | | | | | | |
|---|---|---|---|---|---|---|---|---|---|---|---|---|
| Somalia | 8.3 | 7.1 | 11.2 | 8.0 | 7.3 | 7.9 | 8.1 | 9.0 | 9 | 7.5 | 9 | 7 |
| South Africa | 10.2 | 7.7 | 8.6 | 9.0 | 7.8 | 7.6 | 7.3 | 9.6 | 9 | 8.0 | 10 | 8 |
| Spain | 10.4 | 10.4 | 9.8 | 10.0 | 11.0 | 9.6 | 7.7 | 11.2 | 10 | 11.7 | 11 | 12 |
| Sri Lanka | 11.1 | 7.7 | 11.7 | 10.0 | 8.5 | 11.4 | 10.3 | 11.2 | 10 | 9.7 | 10 | 10 |
| Sudan | 11.4 | 7.5 | 13.4 | 10.0 | 7.7 | 9.8 | 9.4 | 12.6 | 11 | 7.9 | 11 | 8 |
| Suriname | 10.7 | 11.5 | 11.9 | 12.0 | 13.7 | 10.0 | 10.0 | 12.2 | 12 | 13.1 | 12 | 13 |
| Swaziland | 10.2 | 7.1 | 8.0 | 7.0 | 7.1 | 6.1 | 5.2 | 8.1 | 8 | 7.6 | 8 | 8 |
| Sweden | 11.7 | 10.4 | 9.2 | 10.0 | 11.0 | 9.1 | 7.9 | 12.3 | 11 | 11.8 | 11 | 12 |
| Switzerland | 11.0 | 10.9 | 8.8 | 11.0 | 11.9 | 8.4 | 8.1 | 12.1 | 11 | 12.2 | 11 | 12 |
| Syrian Arab Republic | 12.1 | 8.7 | 12.9 | 11.0 | 9.1 | 12.7 | 10.5 | 12.7 | 11 | 9.5 | 11 | 10 |
| Tajikistan | 9.8 | 8.0 | 12.7 | 8.0 | 7.4 | 13.7 | 10.1 | 10.5 | 9 | 7.6 | 9 | 8 |
| Thailand | 10.0 | 8.4 | 10.5 | 10.0 | 8.7 | 11.5 | 10.2 | 9.7 | 11 | 9.5 | 10 | 10 |
| The former Yugoslav Republic of Macedonia | 10.5 | 8.4 | 8.9 | 10.0 | 8.6 | 11.0 | 10.2 | 10.8 | 10 | 9.7 | 10 | 10 |
| Togo | 10.1 | 9.5 | 10.3 | 9.0 | 10.5 | 8.2 | 7.7 | 10.1 | 9 | 10.9 | 9 | 11 |
| Tonga | 10.3 | 7.7 | 10.8 | 10.0 | 7.9 | 10.5 | 9.6 | 10.6 | 9 | 8.0 | 9 | 8 |
| Trinidad and Tobago | 10.8 | 7.2 | 10.7 | 10.0 | 7.6 | 10.6 | 8.6 | 12.0 | 11 | 7.7 | 11 | 8 |
| Tunisia | 10.7 | 10.7 | 11.7 | 10.0 | 9.8 | 9.8 | 10.3 | 11.4 | 11 | 9.5 | 10 | 10 |
| Turkey | 10.8 | 9.1 | 12.0 | 11.0 | 9.9 | 10.1 | 9.3 | 11.7 | 11 | 10.6 | 12 | 11 |
| Turkmenistan | 9.3 | 7.7 | 11.9 | 8.0 | 7.4 | 12.7 | 9.7 | 10.4 | 8 | 7.6 | 9 | 8 |
| Tuvalu | | 6.9 | 10.0 | 9.0 | 7.0 | 9.7 | 8.3 | | 10 | 7.3 | 10 | 7 |
| Uganda | 8.8 | 7.7 | 9.4 | 6.0 | 7.3 | 7.9 | 7.2 | 9.7 | 8 | 8.0 | 9 | 8 |
| Ukraine | 10.2 | 8.9 | 12.0 | 8.0 | 8.8 | 11.5 | 9.4 | 10.0 | 9 | 9.0 | 9 | 9 |
| United Arab Emirates | 11.6 | 7.7 | 12.5 | 11.0 | 7.9 | 11.5 | 10.9 | 12.4 | 11 | 8.3 | 11 | 8 |
| United Kingdom | 10.9 | 9.1 | 8.5 | 10.0 | 9.8 | 9.0 | 8.4 | 11.8 | 11 | 10.2 | 11 | 10 |
| United Republic of Tanzania | 9.3 | 8.0 | 9.6 | 7.0 | 8.1 | 7.9 | 6.8 | 10.0 | 10 | 8.7 | 10 | 9 |
| United States of America | 10.5 | 9.1 | 10.7 | 10.0 | 9.6 | 10.7 | 8.5 | 11.0 | 10 | 10.3 | 10 | 10 |
| Uruguay | 10.2 | 10.0 | 11.4 | 11.0 | 8.8 | 10.9 | 9.9 | 10.4 | 11 | 9.4 | 11 | 9 |
| Uzbekistan | 10.5 | 7.1 | 12.2 | 9.0 | 6.9 | 12.4 | 10.0 | 10.6 | 10 | 7.5 | 10 | 8 |
| Vanuatu | 9.1 | 8.1 | 10.8 | 11.0 | 8.5 | 10.8 | 9.8 | 9.5 | 11 | 9.0 | 10 | 9 |
| Venezuela, Bolivarian Republic of | 10.5 | 8.5 | 12.3 | 10.0 | 8.8 | 11.5 | 10.1 | 10.7 | 11 | 9.1 | 11 | 9 |
| Viet Nam | 10.1 | 8.2 | 11.3 | 11.0 | 8.5 | 10.4 | 9.3 | 10.5 | 11 | 8.6 | 10 | 9 |
| Yemen | 10.3 | 7.8 | 12.7 | 10.0 | 7.9 | 10.2 | 11.5 | 11.0 | 10 | 8.0 | 11 | 8 |
| Zambia | 8.6 | 7.4 | 7.2 | 6.0 | 6.8 | 5.6 | 5.3 | 8.6 | 8 | 8.9 | 9 | 9 |
| Zimbabwe | 11.0 | 8.9 | 7.9 | 6.0 | 7.3 | 5.5 | 4.7 | 9.0 | 9 | 8.8 | 9 | 9 |
| Method | HALE | MODEL | HALE | HALEb | MODEL | HALE | HALE | HALE | HALE | MODEL | HALE | MODEL |
| Year | 1990 | 1990 | 2000 | 2000 | 2000 | 2001 | 2002 | 2010 | 2012 | 2012 | 2013 | 2013 |
| Mean | 10.3 | **8.6** | 11.0 | 9.5 | **8.8** | 10.0 | 8.9 | 10.8 | 10.2 | **9.3** | 10.1 | 9.3 |